\documentclass{article}
\PassOptionsToPackage{sort, numbers, compress}{natbib}
\usepackage[preprint]{neurips_2026} %"Preprint. Work in progress."

\usepackage[utf8]{inputenc} % allow utf-8 input
\usepackage[T1]{fontenc}    % use 8-bit T1 fonts
\usepackage{url}            % simple URL typesetting
\usepackage[breaklinks,colorlinks]{hyperref}
\hypersetup{
  colorlinks=true,
  citecolor={rgb,255:red,0;green,100;blue,224},   % 文献引用颜色 #a1c5f3
  linkcolor={rgb,255:red,224;green,93;blue,166},      % 图表引用颜色 #FF5733
  urlcolor={rgb,255:red,0;green,100;blue,224}        % URL颜色 #0096FF
}
\usepackage{booktabs}       % professional-quality tables
\usepackage{amsfonts}       % blackboard math symbols
\usepackage{amssymb}
\usepackage{graphicx}
\usepackage{subcaption}
\usepackage{multirow}
\usepackage{multicol}
\usepackage[normalem]{ulem}
\usepackage{wrapfig}
\usepackage{amsmath}
\usepackage{nicefrac}       % compact symbols for 1/2, etc.
\usepackage{microtype}      % microtypography
\usepackage[table]{xcolor}  % colors + \rowcolor support

\usepackage{booktabs}
\usepackage{multirow}
\usepackage{makecell}
\usepackage{threeparttable}
\usepackage{array}          % needed for \arraystretch tweaks / column specs
\usepackage{colortbl}       % \rowcolor, \arrayrulecolor
\usepackage{pifont}         % \ding{51}, \ding{55} (check / cross marks)
\usepackage{fontawesome5}   % \faRocket and other small icons for tables
\usepackage{enumitem}

\usepackage[most,skins,theorems]{tcolorbox}
\usepackage{listings}
\newtcolorbox{promptbox}[1][]{
  enhanced, breakable,
  colback=gray!1,      
  colframe=gray!60,    
  coltitle=black,      
  boxrule=2pt,
  arc=10pt,
  left=6pt, right=6pt, top=6pt, bottom=6pt,
  title={#1}, fonttitle=\bfseries,
  attach boxed title to top left={yshift*=-3mm},
  boxed title style={colback=gray!10}
}

\definecolor{citeblue}{RGB}{0,100,224}

\newtcolorbox{templatebox}[1]{
  enhanced,
  breakable,
  colback=citeblue!2,
  colframe=citeblue!35,
  colbacktitle=citeblue!65,
  coltitle=white,
  boxrule=0.9pt,
  arc=6pt,
  outer arc=6pt,
  left=6pt,
  right=6pt,
  top=6pt,
  bottom=6pt,
  title={#1},
  fonttitle=\bfseries,
  boxed title style={
    arc=6pt,
    outer arc=6pt,
    boxrule=0pt
  }
}

\newlength{\loglabelwidth}
\lstdefinestyle{promptplain}{
  language={},
  basicstyle=\ttfamily\footnotesize,
  keywordstyle=\color{black},
  commentstyle=\color{gray},
  stringstyle=\color{black},
  backgroundcolor=\color{gray!5},
  frame=single,
  rulecolor=\color{black!35},
  numbers=none,
  breaklines=true,
  showstringspaces=false
}

\tcbset{
  aibox/.style={
    width=\linewidth,
    top=8pt,
    bottom=4pt,
    colback=inftythink-red!15,
    colframe=inftythink-red,
    colbacktitle=inftythink-red!90!black,
    enhanced,
    center,
    attach boxed title to top left={yshift=-0.1in,xshift=0.15in},
    boxed title style={boxrule=0pt,colframe=white,},
  }
}
\newtcolorbox{AIbox}[2][]{aibox,title=#2,#1}

\title{SaaSBench: Exploring the Boundaries of Coding Agents in Long-Horizon Enterprise SaaS Engineering}

\author{%
\textbf{Qingnan Ren$^{1}$, Shun Zou$^{1,2}$, Shiting Huang$^{1}$, Ziao Zhang$^{1}$, Kou Shi$^{1}$, Zhen Fang$^{1}$,}\\
\textbf{Yiming Zhao$^{1}$, Yu Zeng$^{1}$, Qisheng Su$^{1}$, Lin Chen$^{1}$, Yong Wang$^{2,\dagger}$, Zehui Chen$^{1}$,}\\
\textbf{Xiangxiang Chu$^{2}$, Feng Zhao$^{1,\dagger}$}\\[4pt]
\normalfont $^{1}$University of Science and Technology of China \qquad $^{2}$AMAP, Alibaba Group\\
\normalfont $^{\dagger}$Corresponding authors
}

\begin{document}

\maketitle

\begin{abstract}
\label{abstract}

As autonomous coding agents become capable of handling increasingly long-horizon tasks, they have gradually demonstrated the potential to complete end-to-end software development.
Although existing benchmarks have recently evolved from localized code editing to from-scratch project generation, they remain confined to structurally simplified, single-stack applications. 
Consequently, they fail to capture the heterogeneous environments, full-stack orchestration, and system-level complexity of real enterprise Software as a Service (SaaS) systems, leaving a critical gap in assessing agents under realistic engineering constraints.
To fill this gap, we introduce SaaSBench, the first benchmark designed to explore the boundaries of AI agents in enterprise SaaS engineering.
Spanning 30 complex tasks across 6 SaaS domains with 5,370 validation nodes, it incorporates 8 programming languages, 6 databases, and 13 frameworks to meticulously mirror real-world software heterogeneity.
Furthermore, we design a dependency-aware hybrid evaluation paradigm tailored for complex systems with long horizons and multi-component coupling, enabling fine-grained, reproducible assessment.
Crucially, our extensive experiments reveal a striking insight: the primary bottleneck for state-of-the-art agents is not generating isolated code logic, but successfully configuring and integrating a multi-component system.
Over 95\% of task failures occur before agents even reach deep business logic, with models often falling victim to overconfidence and prematurely halting during foundational system setup, or getting trapped in ineffective debugging loops.
We hope SaaSBench serves as a practical and challenging testbed to drive the evolution of reliable, system-level coding agents. The code is available at \url{https://github.com/ShadeCloak/SaaSbench}.

\end{abstract}

\section{Introduction}
\label{introduction}
With the rapid development of large language models (LLMs)~\cite{anthropic2025opus4sonnet4,anthropic2026claudeopus47,qwen3.6-27b,jiang2026survey,liu2025deepseek}, coding agents have evolved from early tools primarily designed for function completion and localized editing into systems with composite capabilities, including requirement understanding, system design, code generation, environment interaction, and iterative debugging~\cite{dong2025survey,liu2025deepseek,anthropic2025sonnet45,anthropic2025claudecodeagentic,qwen2026qwen36plus}. They are also entering real software development workflows in diverse forms~\cite{gao2025trae,lin2025se,wang2024openhands,cursor2024cursor,yang2024sweagent,huang2025opencoder}. At the same time, coding agents continue to lower the technical barriers to software development, enabling users without development experience to drive the construction of complete software systems from scratch through natural language requirements~\cite{ge2025survey,sapkota2025vibe,sarkar2025vibe}.

\begin{figure}
    \centering
    \includegraphics[width=1\textwidth]{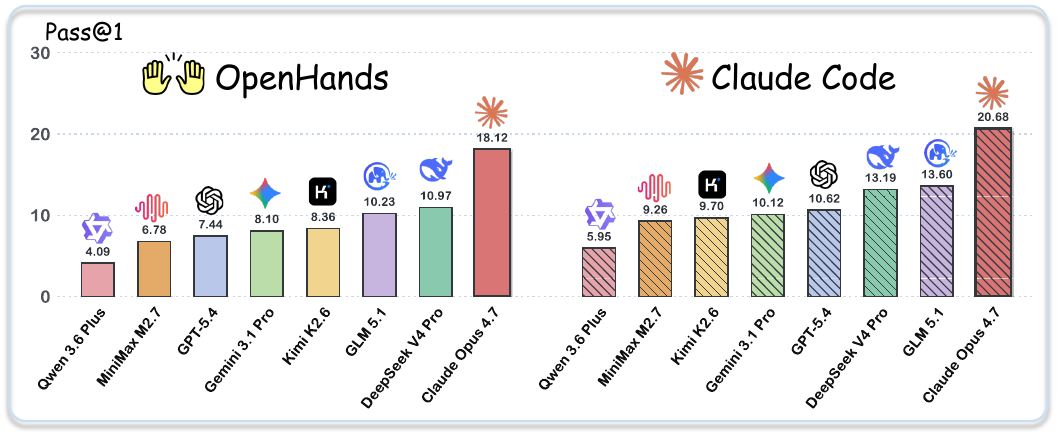}
    \caption{Up-to-Date Leaderboard: Coding agent performance on SaaSBench evaluation tasks.}
    \label{fig:compare}
    \vspace{-6mm}
\end{figure}

Meanwhile, corresponding benchmarks continue to evolve. As shown in Table~\ref{tab:benchmark_comparison}, this trajectory aligns with the expanding capabilities of coding agents. Existing code benchmarks can be broadly divided into two categories. The first mainly focuses on localized and isolated software engineering tasks, such as function-level code generation, patch fixing, and localized modifications within repositories~\cite{chen2021evaluating,austin2021program,hendrycks2021measuring,li2022alphacode,liu2023repobench,jimenez2023swe,deng2025swe,liu2025m2rc}. These benchmarks are better suited for measuring short-horizon and localized engineering behaviors, but they struggle to reflect the holistic capabilities required for end-to-end software development. The second category begins to examine the ability of agents to build complete code repositories or projects from scratch based on natural language requirements, thereby placing higher demands on long-horizon planning, cross-file coordination, and system-level consistency~\cite{li2025prompting,liu2025projecteval,ding2026nl2repobenchlonghorizonrepositorygeneration,peng2026repogenesis,lu2026projdevbench,fu2025automatically,WebGen-Bench}. Although recent project-level and repository-level benchmarks have made progress, they still face three key limitations:

\vspace{-2mm}
\begin{enumerate}[leftmargin=*, itemsep=0.2em, topsep=0.2em, parsep=0pt]
\item \textbf{Lack of real-market grounding.} Existing benchmarks typically define task instances first and then abstract categories from them. As a result, tasks often lack clear market origins, stable product categories, and well-defined business boundaries. This makes it difficult to assess whether an agent truly possesses the ability to build real commercial Software as a Service (SaaS) products.
\item \textbf{Limited system complexity.} Most existing benchmarks operate in software development settings centered on a single language, a single component, or weakly coupled architectures. In contrast, real SaaS system development typically requires the joint design and implementation of the frontend, backend, database, authentication, deployment, and cross-component workflows.
\item \textbf{Insufficient evaluation mechanisms.} Existing evaluations for end-to-end development tasks usually rely on flat end-to-end signals, such as execution outcomes and unit test pass rates. These evaluation methods lack clear definitions and sufficient constraints. They are suitable only for relatively simple software development tasks and fail to characterize prerequisite dependencies, state dependencies, and other constraints in complex real-world business workflows.
\end{enumerate}
\vspace{-2mm}

To address these limitations, we introduce \textbf{SaaSBench}, the first coding agent benchmark systematically designed for real enterprise-level SaaS development scenarios. SaaSBench starts from real software development markets and their open-source product implementations, and constructs the benchmark through a rigorous multi-stage process with strict quality validation. It contains 30 task instances across 6 high-level SaaS domains, covering mainstream SaaS software development scenarios. Each task consists of a long-context product requirements document (PRD), an ambiguity-resolution knowledge base (KB), a standardized runtime environment, and an accompanying DAG-based test suite. This design evaluates whether coding agents can complete the full engineering loop from scratch, including requirement understanding, system implementation, debugging, deployment, and execution. Overall, the PRDs in SaaSBench contain approximately 4,363 lines on average. The benchmark includes 5,370 executable validation nodes and covers 8 programming languages, 6 database types, and 13 frontend and backend development frameworks, reflecting the complexity and diversity of real-world software development.

\definecolor{guibg}{RGB}{248,251,255}     
\definecolor{persbg}{RGB}{248,251,255}    
\definecolor{oursbg}{RGB}{219,234,254}    
\definecolor{oursname}{RGB}{29,78,216}    
\definecolor{oursrule}{RGB}{37,99,235}    
\definecolor{sectcolor}{RGB}{107,114,128} 
\definecolor{cmarkcolor}{RGB}{22,163,74}  
\definecolor{pmarkcolor}{RGB}{126,126,126} 
\definecolor{xmarkcolor}{RGB}{220,38,38}  
\definecolor{nacolor}{RGB}{156,163,175}   

\providecommand{\cmark}{}\providecommand{\pmark}{}\providecommand{\xmark}{}
\renewcommand{\cmark}{\textcolor{cmarkcolor}{\ding{51}}}
\renewcommand{\pmark}{\textcolor{pmarkcolor}{\ding{51}}}
\renewcommand{\xmark}{\textcolor{xmarkcolor}{\ding{55}}}
\providecommand{\namark}{\textcolor{nacolor}{--}}

\providecommand{\rocket}{\faRocket}

\providecommand{\secrow}[2]{%
  \midrule
  \multicolumn{#1}{l}{\small\itshape\color{sectcolor} #2} \\
  \midrule
}
\begin{table*}[t]
\centering
\setlength{\tabcolsep}{3.6pt}
\renewcommand{\arraystretch}{1.22}
\caption{%
  \textbf{Comparison of SaaSBench with representative coding agent
  benchmarks}.
  \cmark~fully incorporated;\enskip
  \pmark~partially incorporated;\enskip
  \xmark~not incorporated;\enskip
  \namark~not applicable.
}
\label{tab:benchmark_comparison}
\resizebox{\textwidth}{!}{%
\begin{tabular}{l ccc ccccc cc}

\toprule[1.3pt]

\multirow{3}{*}[-0.5ex]{\textbf{Benchmark}}
  & \multicolumn{3}{c}{\textbf{Task Realism}}
  & \multicolumn{5}{c}{\textbf{System Complexity}}
  & \multicolumn{2}{c}{\textbf{Evaluation Fidelity}} \\
\cmidrule(lr){2-4} \cmidrule(lr){5-9} \cmidrule(lr){10-11}

  & \makecell{\textbf{From}\\\textbf{Scratch}}
  & \makecell{\textbf{Market-}\\\textbf{Grounded}}
  & \makecell{\textbf{Deployable}\\\textbf{Runtime}}
  & \makecell{\textbf{Multi-}\\\textbf{Language}}
  & \makecell{\textbf{Language}}
  & \makecell{\textbf{Cross-}\\\textbf{Component}}
  & \makecell{\textbf{Workflow}\\\textbf{Depth}}
  & \makecell{\textbf{PRD}\\\textbf{Lines}~\textcolor{oursname}{\rocket}}
  & \makecell{\textbf{Semantic}\\\textbf{Judge}}
  & \makecell{\textbf{Workflow}\\\textbf{Dependency}} \\

\secrow{11}{Snippet-Level Coding Benchmarks}

\rowcolor{guibg}
HumanEval~\cite{chen2021evaluating}
  & \xmark & \xmark & \xmark
  & \xmark & Python & \xmark & \xmark & \namark
  & \xmark & \xmark \\

\rowcolor{guibg}
MBPP~\cite{austin2021program}
  & \xmark & \xmark & \xmark
  & \xmark & Python & \xmark & \xmark & \namark
  & \xmark & \xmark \\

\rowcolor{guibg}
APPS~\cite{hendrycks2021measuring}
  & \xmark & \xmark & \xmark
  & \xmark & Python & \xmark & \xmark & \namark
  & \xmark & \xmark \\

\rowcolor{guibg}
DS-1000~\cite{lai2023ds}
  & \xmark & \xmark & \xmark
  & \xmark & Python & \xmark & \xmark & \namark
  & \xmark & \xmark \\

\rowcolor{guibg}
CodeContests~\cite{li2022alphacode}
  & \xmark & \xmark & \xmark
  & \cmark & Py / Ja / C\texttt{++} & \xmark & \xmark & \namark
  & \xmark & \xmark \\

\rowcolor{guibg}
EvoCodeBench~\cite{EvoCodeBench}
  & \xmark & \xmark & \xmark
  & \xmark & Python & \xmark & \xmark & \namark
  & \xmark & \xmark \\

\rowcolor{guibg}
SWE-Bench~\cite{jimenez2023swe}
  & \xmark & \xmark & \xmark
  & \xmark & Python & \xmark & \xmark & \namark
  & \xmark & \xmark \\

\rowcolor{guibg}
GitTaskBench~\cite{ni2026gittaskbench}
  & \xmark & \pmark & \xmark
  & \xmark & Python & \xmark & \pmark & \namark
  & \xmark & \xmark \\

\addlinespace[2pt]

\secrow{11}{Repository- \& Project-Level Coding Benchmarks}

\rowcolor{persbg}
ProjectEval~\cite{liu2025projecteval}
  & \pmark & \xmark & \xmark
  & \xmark & Python & \xmark & \xmark & 34.45
  & \xmark & \xmark \\

\rowcolor{persbg}
NL2Repo-Bench~\cite{ding2026nl2repobenchlonghorizonrepositorygeneration}
  & \cmark & \xmark & \xmark
  & \xmark & Python & \xmark & \xmark & 2452.64
  & \xmark & \xmark \\

\rowcolor{persbg}
RepoGenesis~\cite{peng2026repogenesis}
  & \cmark & \xmark & \cmark
  & \cmark & Py / Ja & \cmark & \xmark & 210.17
  & \xmark & \xmark \\

\rowcolor{persbg}
PRDBench~\cite{fu2025automatically}
  & \cmark & \xmark & \xmark
  & \xmark & Python & \xmark & \xmark & 105.22
  & \cmark & \xmark \\

\rowcolor{persbg}
ProjDevBench~\citep{lu2026projdevbench}
  & \pmark & \xmark & \xmark
  & \xmark & C\texttt{++} & \xmark & \xmark & 283.85
  & \cmark & \xmark \\

\midrule[0.8pt]

\rowcolor{oursbg}
\llap{\textcolor{oursrule}{\rule[-6.5pt]{2pt}{18pt}}\hspace{4pt}}%
\textcolor{oursname}{\textbf{SaaSBench (Ours)}}
  & \cmark & \cmark & \cmark
  & \cmark & \makecell{Py / Ja / Go \ldots\\(8 langs.)}
  & \cmark & \cmark
  & \textbf{4362.7}\,\textcolor{oursname}{\rocket}
  & \cmark & \cmark \\

\bottomrule[1.3pt]

\end{tabular}
}%
\vspace{-4mm}

\end{table*}

In addition, we design a dependency-aware hybrid evaluation paradigm for long-horizon and highly interactive end-to-end system development tasks. The paradigm centers on a directed acyclic graph (DAG), where each validation node is compiled into a linear checking chain composed of executable primitives. Through prerequisite dependency gating, failure propagation control, and three scoring mechanisms, namely \emph{binary}, \emph{weighted}, and \emph{llm-as-judge}, it enables reproducible and objective evaluation. The validation nodes cover six capability dimensions: deployment availability, data modeling, API contract consistency, business logic correctness, access control, and engineering quality. These dimensions systematically cover the key engineering aspects that must be verified across the lifecycle of real software development.

As shown in Figure~\ref{fig:compare}, our experiments reveal that even state-of-the-art coding agents exhibit substantial capability gaps on SaaSBench, highlighting their limitations in long-horizon task planning and cross-component coordination. These findings provide a foundation for further improving the capabilities of future coding agents. Our main contributions are summarized as follows: 
\begin{itemize}[leftmargin=*,topsep=1pt,itemsep=0.4pt]
\item We introduce \textbf{SaaSBench}, the first benchmark platform designed to evaluate the ability of coding agents to generate and deploy enterprise-level SaaS systems from scratch. It covers mainstream software development markets.
\item We design a dependency-aware hybrid evaluation paradigm for end-to-end complex system development tasks. It provides a reproducible and reliable evaluation mechanism and comprehensively covers the key engineering dimensions in the lifecycle of real software development.
\item We systematically evaluate a broad range of agents and models on SaaSBench. The results show that even the strongest current agents still face severe challenges in enterprise-level SaaS development.
\end{itemize}

\section{Related Work}
\paragraph{Autonomous Coding Agents.} As the coding capabilities of LLMs continue to improve~\cite{anthropic2026claudeopus47,qwen3.6-27b,jiang2026survey}, coding agents have become essential tools in everyday software development. Modern coding agents can be broadly divided into two categories. The first consists of \emph{IDE-integrated assistants}, such as Cursor, Claude Code, and Codex, which evolve from context-aware code completion toward cross-file modification and repository-level iterative assistance~\cite{github2021copilot,cursor2024cursor,anthropic2025claudecodeagentic,openai2025codexcli}. The second consists of \emph{autonomy-oriented frameworks}, such as OpenHands, Qwen-Agent, and SWE-agent, which incorporate the terminal, file system, and runtime environment into a unified agent loop to support longer-horizon planning, implementation, and debugging~\cite{wang2024openhands,yang2024sweagent,hong2024metagpt}. Despite differences in interaction interfaces and product forms, the two categories exhibit a common trend: they integrate terminal access, script execution, dependency installation, and test feedback into the standard workflow, enabling agents to handle end-to-end software engineering tasks with complex dependencies and long feedback loops.

\paragraph{Code-Centric Agent Benchmarks.} Benchmarks for coding agents have continuously expanded in coverage. Early works such as HumanEval, MBPP, APPS, and CodeContests mainly evaluate function-level code generation in isolated settings~\cite{chen2021evaluating,austin2021program,hendrycks2021measuring,li2022alphacode,xu2025swecompassunifiedevaluationagentic,CodeContests+,BigCodeBench}. Later, RepoBench and SWE-Bench extend evaluation to real code repositories, requiring agents to perform completion, editing, and issue fixing across multiple files~\cite{liu2023repobench,jimenez2023swe,deng2025swe,ni2026gittaskbench,he2025swe,pmlr-v267-miserendino25a,liu2025m2rc}. However, these settings remain largely incremental and primarily measure localized, short-horizon engineering capabilities.

A recent line of work further requires agents to build complete code repositories or projects from scratch. 

NL2Repo-Bench~\cite{ding2026nl2repobenchlonghorizonrepositorygeneration} generates complete Python projects from specification documents. PRDBench~\cite{fu2025automatically} uses product requirements documents (PRDs) as the core input. RepoGenesis~\cite{peng2026repogenesis} targets repository-level web microservice generation. ProjDevBench~\cite{lu2026projdevbench} further incorporates Online Judge diagnostic signals and LLM-based code review. Although these works make progress in repository-level and project-level evaluation, a substantial gap remains between their settings and real enterprise-level SaaS system development. They also lack stable automated evaluation protocols for highly interactive and multi-dependency systems, which is the gap that SaaSBench aims to fill.

\begin{figure}
    \centering
    % \vspace{-3mm}
    \includegraphics[width=1\textwidth]{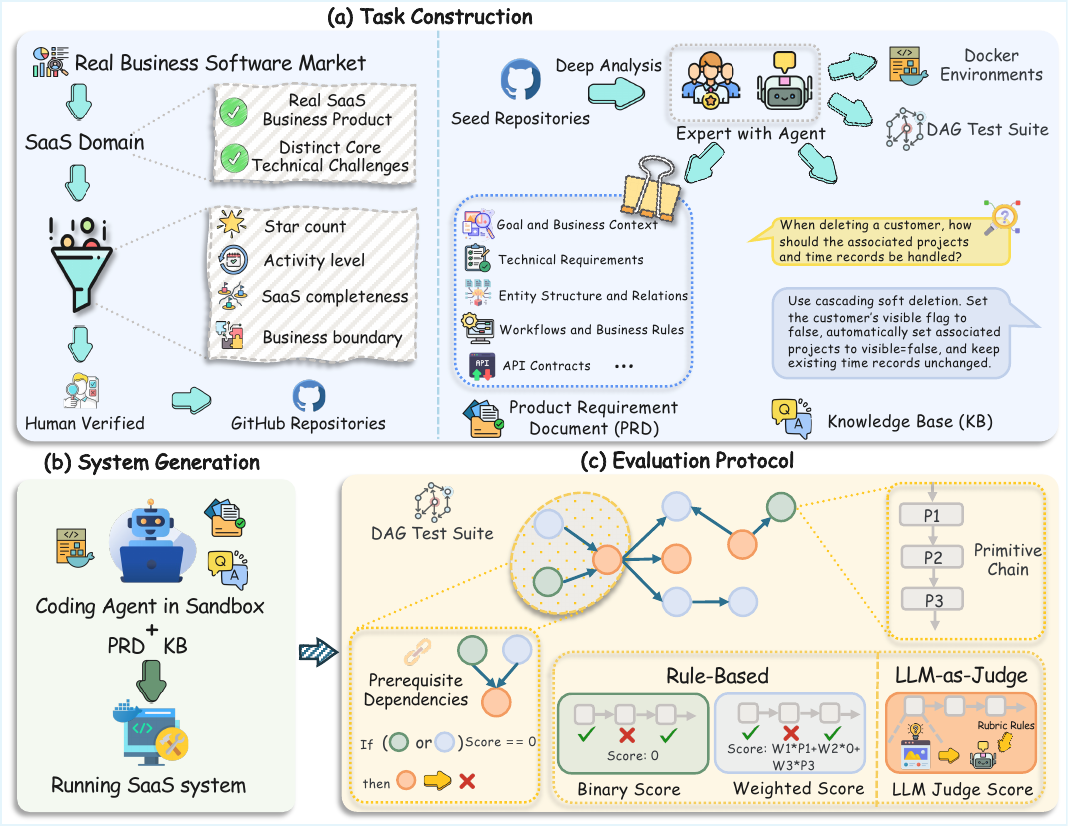}
    \caption{Overview of SaaSBench. The benchmark is grounded in real software development markets and constructs tasks through a multi-stage human-agent collaborative process. Evaluation is conducted with a reproducible dependency-aware hybrid evaluation paradigm.}
    \label{fig:overview}
    \vspace{-5mm}
\end{figure}

\section{SaaSBench}
As shown in Figure~\ref{fig:overview}, the construction of SaaSBench is carried out through collaboration between experienced doctoral researchers and Cursor~\cite{cursor2024cursor}. Building a single task requires a multi-stage systematic workflow, including candidate repository auditing, PRD writing, KB organization, standardized container environment preparation, DAG test-suite implementation, and strict quality validation. The detailed construction workflow is presented in the following subsections.

\subsection{Benchmark Construction}
\label{benchmark construction}
\noindent\textbf{SaaS Domain Definition and Seed Repository Selection.}
SaaSBench defines candidate domains from real software development markets. Specifically, we refer to industry taxonomies, publicly available commercial product landscapes, and consultations with domain experts. We retain only domains that satisfy two conditions. First, the domain corresponds to stable commercial SaaS use cases and identifiable product forms. Second, the core technical challenges introduced by the domain are not substantially redundant with those of other selected domains. The resulting task space is therefore clearly grounded in real markets while preserving diversity in engineering patterns.

For each selected domain, we further select corresponding seed repositories. Candidate repositories must satisfy the following requirements. They need to show signals of continuous maintenance and community activity, provide a complete SaaS system form, and maintain a clear primary business boundary, meaning that each repository mainly serves one interpretable business domain. Annotators then conduct cold-start validation on the candidate repositories, requiring each repository to be independently built, successfully launched, and verified through basic smoke tests. Detailed descriptions of the domains and repositories are provided in Appendix~\ref{appendix-Domain Selection Principles} and \ref{appendix-Repo Statistics}.

\noindent\textbf{PRD Construction.}
After determining the seed repositories, we construct PRDs through a rigorous workflow. First, annotators and agents analyze each repository in depth, systematically examining its code structure, configuration files, route definitions, data models, existing tests, and key business logic. Based on this analysis, we generate comprehensive long-context PRDs. Unlike most benchmarks that retain only short problem descriptions or feature lists, the PRDs in SaaSBench preserve as much key information as possible for system-level development, including technical requirements, complete data models, core business workflows, API contracts, permission policies, boundary rules, deployment constraints, and build steps. This makes them closer to the long-document requirement inputs used in real enterprise development. We ensure that each PRD provides complete coverage of all major aspects of the corresponding repository.

\noindent\textbf{KB Construction and Environment Building.}
In real-world development, clients often provide further revisions and detailed feedback based on an initial product prototype. Similarly, a PRD alone is insufficient to express all evaluation-sensitive details. We therefore further construct an ambiguity-resolution KB. Each KB record corresponds to a behavioral detail that affects correctness but is difficult to express stably in natural language requirements, such as default pagination rules, deletion semantics, or fallback logic. This reduces ambiguity in requirement descriptions and helps ensure the stability and auditability of evaluation. In addition, we build a standardized runtime environment for each task. The environment artifacts are containerized and preinstall the required system packages, system dependencies, database services, port mappings, and environment variables for the corresponding task.

\subsection{DAG Evaluation Protocol}
\noindent\textbf{Motivation.}
For end-to-end enterprise-level SaaS development, a conventional list of unit tests is insufficient for reliable evaluation. First, failures in foundational capabilities often introduce secondary noise into many downstream tests, obscuring the true bottlenecks. Second, if evaluation relies only on shallow signals, such as file existence or basic CRUD functionality, an agent may receive a high score even when it fails to correctly implement key business semantics. The fundamental reason is that multi-user interactions, multi-model data operations, and cross-module business workflows in real SaaS systems are not independent. Instead, they form long-horizon interaction processes built on shared application states and explicit prerequisite dependencies.

\noindent\textbf{Definition of the DAG-based Hybrid Evaluation Paradigm.}
Based on these observations, we organize the evaluation paradigm as a DAG \(G=(V,E)\). Each node \(v \in V\) corresponds to an independently scored validation unit, and each edge \(e \in E\) explicitly represents a prerequisite dependency between nodes. Each node contains a primitive chain composed of basic validation primitives executed in sequence. The executable checks include HTTP requests, authentication login, and rubric-based LLM judgment, among others, as detailed in Table~\ref{tab:primitives}. Node scoring falls into three categories. \emph{binary} is used for scenarios that must be fully correct, such as permission gating and security constraints. \emph{weighted} is used for scenarios that allow partial completion, such as multi-step CRUD workflows. \emph{llm-as-judge} is used only when deterministic assertions cannot adequately characterize the target, such as the reasonableness of page layout. In addition, we assign each evaluation node to one of six engineering capability dimensions: Deploy, Data, API, Logic, AuthZ, and Quality, enabling comprehensive evaluation of software engineering dimensions.

\noindent\textbf{DAG Test Suite Construction.}
The DAG is not constructed by manually listing test items in an arbitrary manner. Instead, it follows a comprehensive and complete definition and is systematically compiled from the task artifacts. First, annotators collaborate with agents to scan the PRD and map each verifiable requirement to a candidate node. Next, any assertion involving potential ambiguity must be aligned with the KB. Finally, each node must be compiled into an executable linear chain of primitives and assigned prerequisite dependencies that reflect real business workflows. Detailed definitions are provided in Appendix~\ref{appendix-primitive-taxonomy}.

\noindent\textbf{Evaluation Pipeline.}

As shown in Figure~\ref{fig:overview}, during evaluation, the agent receives two inputs, the PRD and the KB, together with a carefully designed prompt, as detailed in Appendix~\ref{appendix-full-prompts}, and runs in an isolated, pre-built Docker environment. Within the specified workspace of this environment, the agent is granted full autonomy to build and deploy a runnable and accessible SaaS system from scratch, without any human intervention. The evaluation system then topologically sorts the DAG test suite corresponding to the task and executes the evaluation nodes on the running system one by one in dependency order. If any prerequisite dependency of a node is not satisfied, the node is not simply marked as a direct failure. Instead, it is marked as \emph{Skipped dependency}, which prevents foundational errors from being repeatedly penalized across all downstream nodes.
 
\begin{figure}
    \centering
    \includegraphics[width=1\textwidth]{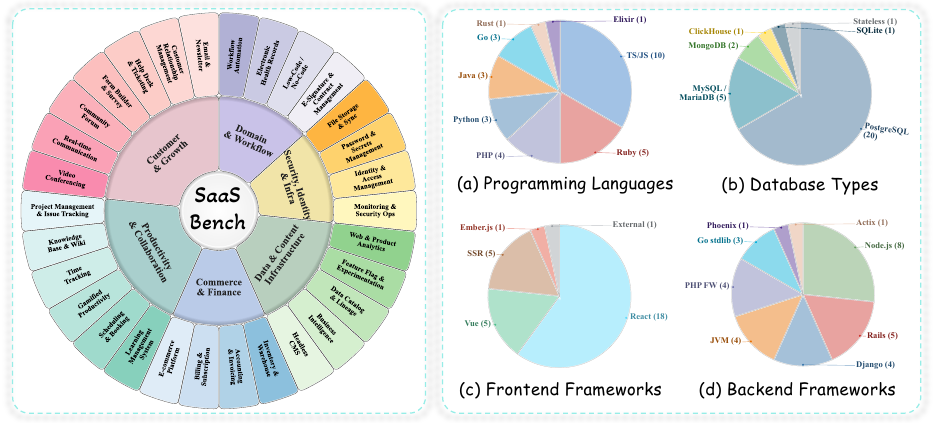}
    \caption{Statistical overview of SaaSBench. Left: SaaSBench includes six key SaaS domains and 30 fine-grained categories, covering mainstream software development markets. Right: Distribution of tasks across programming languages, database types, and frontend and backend frameworks.}

    \label{fig:data}
    \vspace{-6mm}
\end{figure}

\subsection{Task Quality Validation}
\noindent\textbf{PRD Alignment Verification.}
To ensure the completeness and accuracy of each PRD, we introduce an independent review and revision loop. After the initial PRD is completed, two additional annotators inspect the seed repository and verify the PRD with a structured checklist. For each missing, inconsistent, or underspecified requirement, the reviewers record a revision item and return it to the PRD author for refinement. The revised PRD is checked again until the reviewers confirm that it covers the key capabilities of the repository. This process reduces requirement omissions and hallucinated requirements, and improves the alignment between each task and the corresponding executable SaaS system.

\noindent\textbf{Test-Suite Quality Assurance.}
To avoid subtle errors in the test suite, such as incorrect assertions or fragile chains of atomic capability calls, we conduct strict quality validation for each task. Specifically, we deploy the upstream source code of the seed repository in the same standardized runtime environment used for evaluation, and require the reference implementation to pass the full test suite. For \emph{llm-as-judge} nodes, we allow bounded variance. Only tasks that pass this validation are included in the benchmark. Tasks that fail this gate are revised until they converge.

\subsection{Benchmark Statistics}
SaaSBench exhibits clear characteristics of real SaaS development in terms of task coverage, technology-stack diversity, and system complexity. As shown in Figure~\ref{fig:data}, SaaSBench contains 30 tasks, covering 6 high-level domains and 30 fine-grained SaaS categories. Each task typically includes a frontend interface, backend APIs, persistent data models, a role-based permission system, and deployment configurations, forming a system structure that is clearly distinct from function-level, patch-level, and toy project-level or repository-level benchmarks.

Overall, the PRDs in SaaSBench contain 4,363 lines on average. The benchmark includes 5,370 executable validation nodes connected by 6,167 prerequisite dependency edges, and covers 8 programming languages, 6 types of database systems, 5 types of frontend frameworks, and 8 types of backend frameworks. More fine-grained per-task statistics and technology-stack distributions are provided in Appendix~\ref{appendix-Repo Statistics}.

\section{Experiments}
\label{experiment}

\definecolor{agentbg}{HTML}{F6F8FC}
\definecolor{agentname}{HTML}{1E3A8A}
\definecolor{dashcolor}{HTML}{9CA3AF}
\definecolor{headerblue}{HTML}{5B84D7}

\providecommand{\dash}{\textcolor{dashcolor}{--}}

\newcolumntype{L}[1]{>{\raggedright\arraybackslash}p{#1}}
\newcolumntype{C}[1]{>{\centering\arraybackslash}p{#1}}

\newcommand{\openhandslogo}{%
  \raisebox{-0.12ex}{\includegraphics[height=0.90em]{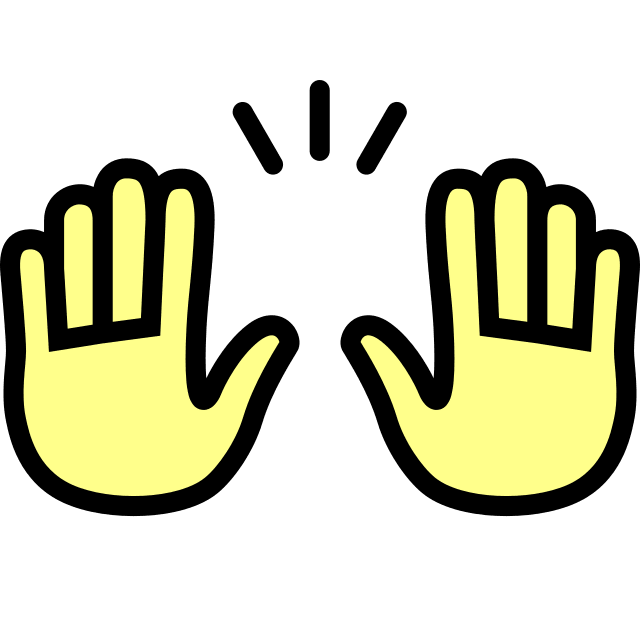}}%
}

\newcommand{\claudecodelogo}{%
  \raisebox{-0.12ex}{\includegraphics[height=0.90em]{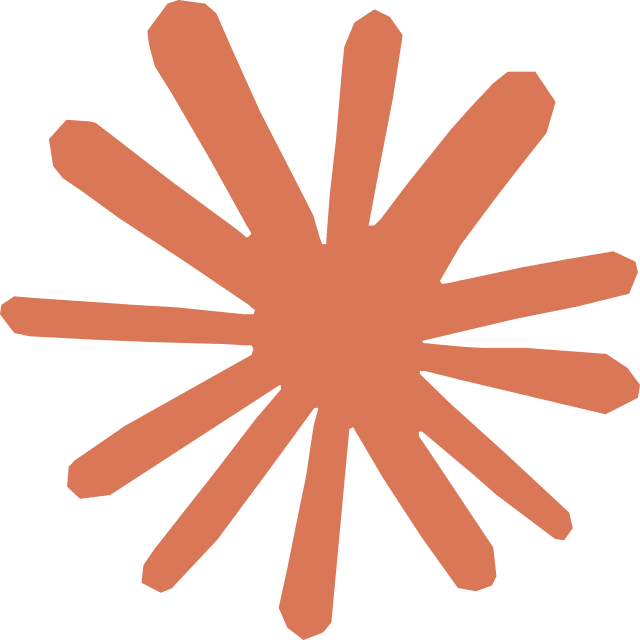}}%
}

\newcommand{\agentrow}[2]{%
  \specialrule{0.35pt}{1.5pt}{0pt}
  \rowcolor{agentbg}
  \multicolumn{9}{l}{%
    \rule{0pt}{2.5ex}%
    \textcolor{agentname}{\textbf{\textit{#1}}}%
    \hspace{0.45em}#2%
  } \\
  \specialrule{0.35pt}{0pt}{2.2pt}
}

\newcommand{\resultrow}[9]{%
  \strut #1
  & #2 & #3
  & #4 & #5 & #6 & #7 & #8 & #9
  \\[0.1pt]
}

\renewcommand{\arraystretch}{1.04}
\setlength{\tabcolsep}{2.6pt}

\begin{table*}[t]
\centering
\scriptsize
\caption{
Main results on SaaSBench across different agent--model configurations.
We report the overall Pass@1 and Node Coverage, together with scores over the six SaaS domains.
Within each coding agent block, \textbf{bold} numbers indicate the best performance, and \underline{underlined} numbers indicate the second-best performance.
}
\label{tab:domain-results}

\resizebox{\textwidth}{!}{%
\begin{tabular}{
@{}
L{2.45cm}
C{0.95cm}
C{1.05cm}
*{6}{C{0.78cm}}
@{}
}
\toprule[0.8pt]

\multirow{2}{*}{\textbf{Coding Agent}}
&
\multicolumn{2}{c}{\textbf{Overall}}
&
\multicolumn{6}{c}{\textbf{SaaS Domain}}
\\

\arrayrulecolor{headerblue}
\cmidrule(lr){2-3}
\cmidrule(lr){4-9}
\arrayrulecolor{black}

&
\textbf{Pass@1}
&
\textbf{Node Cov.}
&
\textbf{CG}
&
\textbf{PC}
&
\textbf{CF}
&
\textbf{DCI}
&
\textbf{SI}
&
\textbf{DW}
\\

\agentrow{OpenHands}{\openhandslogo}
\resultrow{Qwen 3.6 Plus}
{4.09}{5.85}{4.19}{3.42}{2.10}{2.43}{3.70}{9.38}

\resultrow{Kimi K2.6}
{8.36}{8.94}{5.78}{6.97}{2.60}{5.79}{17.45}{14.88}

\resultrow{MiniMax M2.7}
{6.78}{6.43}{3.59}{5.87}{1.98}{\textbf{13.22}}{3.28}{13.98}

\resultrow{GPT-5.4}
{7.44}{8.83}{7.63}{2.23}{0.98}{6.34}{14.25}{15.96}

\resultrow{Gemini 3.1 Pro}
{8.10}{8.91}{1.94}{3.37}{2.30}{8.46}{\textbf{25.25}}{14.20}

\resultrow{DeepSeek V4 Pro}
{\underline{10.97}}{10.55}{4.29}{\textbf{9.97}}{\underline{4.35}}{10.37}{21.93}{\textbf{20.60}}

\resultrow{GLM 5.1}
{10.23}{\underline{11.21}}{\underline{14.03}}{7.02}{3.97}{\underline{10.65}}{16.22}{8.12}

\resultrow{Claude Opus 4.7}
{\textbf{18.12}}{\textbf{18.24}}{\textbf{36.16}}{\underline{8.62}}{\textbf{4.62}}{7.69}{\underline{24.90}}{\underline{20.55}}

\agentrow{Claude Code}{\claudecodelogo}
\resultrow{Qwen 3.6 Plus}
{5.95}{6.10}{6.85}{7.12}{2.92}{6.48}{0.53}{10.39}

\resultrow{Kimi K2.6}
{9.70}{9.70}{4.49}{\underline{9.90}}{0.48}{10.51}{22.07}{\underline{14.38}}

\resultrow{MiniMax M2.7}
{9.26}{10.62}{11.07}{5.63}{4.92}{8.46}{17.57}{8.56}

\resultrow{GPT-5.4}
{10.62}{9.81}{11.90}{4.17}{2.40}{\underline{14.50}}{20.62}{11.41}

\resultrow{Gemini 3.1 Pro}
{10.12}{10.12}{11.29}{4.28}{1.47}{9.97}{\underline{30.15}}{5.60}

\resultrow{DeepSeek V4 Pro}
{13.19}{12.13}{\underline{15.96}}{8.98}{4.48}{12.78}{22.90}{14.16}

\resultrow{GLM 5.1}
{\underline{13.60}}{\underline{15.28}}{11.08}{\textbf{10.57}}{\underline{6.75}}{13.91}{25.93}{\textbf{16.73}}

\resultrow{Claude Opus 4.7}
{\textbf{20.68}}{\textbf{18.50}}{\textbf{21.51}}{9.30}{\textbf{15.65}}{\textbf{37.14}}{\textbf{34.35}}{7.06}

\bottomrule[0.8pt]
\end{tabular}%
}
\vspace{-6mm}
\end{table*}

\subsection{Experimental Settings}
\label{sec:exp_settings}

\noindent\textbf{Evaluated Agents and LLM Backends.}
We evaluate eight state-of-the-art open-source and closed-source large models, including GPT-5.4~\cite{openai2026gpt54}, Gemini~3.1~Pro~\cite{google2026gemini31propreview}, Claude~Opus~4.7~\cite{anthropic2026claudeopus47}, Kimi~K2.6~\cite{moonshotai2026kimik26}, Qwen~3.6~Plus~\cite{qwen2026qwen36plus}, DeepSeek~V4 Pro~\cite{deepseekai2026deepseekv4}, GLM~5.1~\cite{zai2026glm51}, and MiniMax~M2.7~\cite{minimax2026m27}. We integrate these models into two representative coding agent frameworks: OpenHands~\cite{wang2024openhands} and Claude Code~\cite{anthropic2025claudecodeagentic}. For \emph{llm-as-judge} nodes, we use Claude Sonnet 4.5~\cite{anthropic2025sonnet45} as the rubric judge and set \emph{temperature = 0} to maximize reproducibility. More detailed experimental settings, agent framework descriptions, and model information are provided in Appendix~\ref{appendix-evaluated-agents}.

\noindent\textbf{Evaluation Metrics.}

We mainly report the average pass@1 across all tasks, as well as performance across the six SaaS domains: Customer \& Growth (CG), Productivity \& Collaboration (PC), Commerce \& Finance (CF), Data \& Content Infrastructure (DCI), Security, Identity \& Infrastructure (SI), and Domain \& Workflow Platforms (DW). We also report the node-coverage rate, defined as the proportion of validation nodes that reach the \emph{Passed} state across all tasks. Detailed metric definitions are provided in Appendix~\ref{appendix-metric-detail}.

\subsection{Main Results}
\label{sec:main_results}
\noindent\textbf{SaaSBench is a Highly Challenging Benchmark.}

As shown in Table~\ref{tab:domain-results} and Figure~\ref{fig:compare}, SaaSBench poses a substantial challenge to current coding agents. The best result is only 20.68\%, achieved by Claude Opus 4.7 under Claude Code. On average, Claude Code achieves a performance of 11.64\%, while OpenHands achieves 9.26\%. These results show that current LLM-based coding agents still struggle to reliably generate complete enterprise-level SaaS systems from scratch based on natural language requirements, and that end-to-end SaaS construction remains far from solved. Unlike prior toy-level project or repository generation from scratch, SaaSBench requires agents to jointly handle long-context requirement understanding, multi-step task planning, cross-component implementation, persistent data modeling, permission logic, and deployment-level execution.

\noindent\textbf{Performance Varies across Task Domain.}

We further break down the evaluation results by the six high-level SaaS domains. Overall, models perform relatively better on SI and DCI, while their performance is consistently weaker on CF and PC. This difference suggests that current coding agents can more easily handle infrastructure-oriented tasks with clear structures and well-defined interface boundaries. However, they still face substantial challenges in scenarios involving complex business semantics, long-horizon state dependencies, and multi-user collaborative behavior. In particular, billing, order, and financial-state consistency in CF tasks, as well as calendar interactions and shared-state management in PC tasks, often require agents to maintain cross-component consistency among the frontend, backend, database, and permission logic. These results further characterize the capability boundaries of current models across different SaaS product forms.

\section{Fine-grained Analysis}

\definecolor{mainagentbg}{HTML}{f6f8fc}
\definecolor{mainagentname}{HTML}{1E3A8A}
\definecolor{maindashcolor}{HTML}{9CA3AF}
\definecolor{mainheaderblue}{HTML}{5B84D7}

\providecommand{\dash}{\textcolor{maindashcolor}{--}}

\providecommand{\mainagentrow}{}
\renewcommand{\mainagentrow}[2]{%
  \specialrule{0.35pt}{0pt}{0pt}
  \rowcolor{mainagentbg}
  \multicolumn{7}{l}{%
    \rule{0pt}{2.3ex}%
    \textcolor{mainagentname}{\textbf{\textit{#1}}}%
    \hspace{0.45em}#2%
  } \\
  \specialrule{0.35pt}{0pt}{0.8pt}
}

\providecommand{\mainresultrow}{}
\renewcommand{\mainresultrow}[7]{%
\strut #1
& #2 & #3 & #4 & #5 & #6 & #7
\\[-0.3pt]
}

\begin{wraptable}[23]{r}{0.57\textwidth}
\vspace{-4.5mm}
\centering
\scriptsize
\caption{
Engineering capability dimensions results on SaaSBench.
We report per-category scores for Deploy, Data, API, Logic, AuthZ, and Quality.
}
\label{tab:main-results2}

\renewcommand{\arraystretch}{1.05}
\setlength{\tabcolsep}{2.8pt}

\resizebox{\linewidth}{!}{%
\begin{tabular}{
@{}
>{\raggedright\arraybackslash}p{1.8cm}
*{6}{>{\centering\arraybackslash}p{0.70cm}}
@{}
}

\toprule[0.8pt]

\multirow{2}{*}{\textbf{Coding Agent}}
&
\multicolumn{6}{c}{\textbf{Capability dimensions score}}
\\

\arrayrulecolor{mainheaderblue}
\cmidrule(lr){2-7}
\arrayrulecolor{black}

&
\textbf{Deploy}
&
\textbf{Data}
&
\textbf{API}
&
\textbf{Logic}
&
\textbf{AuthZ}
&
\textbf{Quality}
\\

\mainagentrow{OpenHands}{\openhandslogo}
\mainresultrow{Qwen 3.6 Plus}
{14.12}{8.48}{1.06}{1.02}{3.03}{1.46}

\mainresultrow{Kimi K2.6}
{18.68}{\underline{11.29}}{5.39}{5.85}{7.66}{2.28}

\mainresultrow{MiniMax M2.7}
{16.75}{7.85}{2.85}{2.64}{4.08}{0.93}

\mainresultrow{GPT-5.4}
{\underline{19.31}}{7.34}{4.33}{5.97}{6.81}{1.76}

\mainresultrow{Gemini 3.1 Pro}
{12.81}{8.37}{6.53}{6.03}{\underline{8.24}}{\textbf{3.59}}

\mainresultrow{DeepSeek V4 Pro}
{18.18}{9.67}{\textbf{6.84}}{\underline{6.13}}{8.11}{\underline{3.02}}

\mainresultrow{GLM 5.1}
{18.54}{8.04}{3.36}{4.11}{6.34}{2.18}

\mainresultrow{Claude Opus 4.7}
{\textbf{19.46}}{\textbf{12.40}}{\underline{6.72}}{\textbf{6.87}}{\textbf{9.43}}{2.99}

\mainagentrow{Claude Code}{\claudecodelogo}
\mainresultrow{Qwen 3.6 Plus}
{13.07}{5.97}{0.27}{0.69}{2.46}{0.54}

\mainresultrow{Kimi K2.6}
{11.57}{7.08}{3.96}{4.64}{5.35}{1.18}

\mainresultrow{MiniMax M2.7}
{18.03}{11.92}{3.74}{3.80}{6.18}{1.79}

\mainresultrow{GPT-5.4}
{19.75}{9.27}{3.82}{5.29}{7.48}{1.80}

\mainresultrow{Gemini 3.1 Pro}
{17.69}{10.84}{6.12}{6.77}{\underline{7.98}}{2.32}

\mainresultrow{DeepSeek V4 Pro}
{\underline{21.31}}{10.36}{4.25}{5.23}{6.77}{2.00}

\mainresultrow{GLM 5.1}
{19.25}{\underline{12.88}}{\underline{6.37}}{\underline{7.81}}{7.79}{\textbf{3.02}}

\mainresultrow{Claude Opus 4.7}
{\textbf{22.76}}{\textbf{14.57}}{\textbf{7.92}}{\textbf{8.79}}{\textbf{11.55}}{\underline{2.66}}

\bottomrule[0.8pt]

\end{tabular}%
}

\vspace{-1mm}
\end{wraptable}
\subsection{Performance by Engineering Capability Dimension}
We further report the evaluation results of agents across six engineering capability dimensions: Deploy, Data, API, Logic, AuthZ, and Quality. Detailed definitions are provided in Appendix~\ref{appendix-backbone-mapping}. As shown in Table~\ref{tab:main-results2}, failures are not uniformly distributed. Deploy is usually the highest-scoring dimension, suggesting that current coding agents already possess some ability in service startup and basic runtime environment configuration. Data remains at an intermediate level, while API, Logic, and AuthZ obtain lower scores. This indicates that agents still face clear difficulties in interface contract consistency, persistent state modeling, business state transitions, and role-based access control. The most prominent bottleneck lies in the Quality dimension, where scores are substantially lower than those of the other engineering capability dimensions. This indicates that even when agents can generate services that start successfully and implement some local functionality, they still lack sufficient code organization, frontend rendering quality, edge-case handling, and overall engineering robustness. This trend is qualitatively associated with higher structural and interaction complexity, and provides guidance for improving future coding agents.

\subsection{Agent Frameworks}
\begin{figure}
    \centering
    \includegraphics[width=1\textwidth]{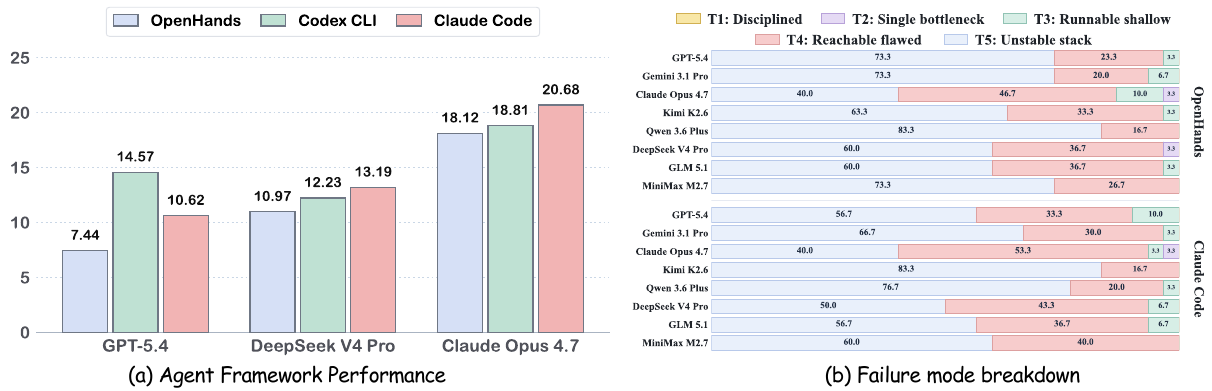}
    \vspace{-4mm}
    \caption{\textbf{Left:} Performance analysis of agent frameworks. \textbf{Right:} We classify capability units into five execution trajectories. T4 and T5
account for 95.6\% of all units, showing that most failures occur before agents
reach deep business logic. See Appendix~\ref{app:failure-mode-taxonomy} for
definitions.}
    \label{fig:analy}
    \vspace{-6mm}
\end{figure}

To examine the impact of agent frameworks beyond the underlying model, we evaluate GPT-5.4 and Claude Opus 4.7 under three agent frameworks: OpenHands, Claude Code, and Codex CLI. As shown in Figure~\ref{fig:analy} (a), the same underlying model can exhibit substantial performance differences across different agent frameworks. Commercial IDE-based frameworks often outperform open-source agent frameworks. This gap mainly stems from how each framework manages tool calls, context, and execution feedback. Commercial IDE-oriented frameworks are usually more tightly integrated with file editing, terminal execution, diagnostic information, and intermediate project states, thereby reducing the burden on the model to track workspace changes and recover from failed commands. In contrast, more open frameworks often require the model to bear greater coordination costs. In long-horizon SaaS tasks, these differences gradually accumulate and lead to clear gaps in deployment stability, dependency repair, schema consistency, and error recovery. This highlights that SaaSBench does not evaluate an isolated LLM, but rather a coupled system composed of the model, tool interfaces, execution loop, environment feedback mechanism, memory mechanism, and error recovery strategy.

\subsection{Interaction Turns and Performance}

\definecolor{mainheaderblue}{HTML}{5B84D7}
\definecolor{interactionheadbg}{HTML}{EEF4FF}

\begin{wraptable}[10]{r}{0.4\textwidth}
\vspace{-5mm}
\centering
\scriptsize
\caption{
Interaction behavior and execution cost of OpenHands on SaaSBench.
}
\label{tab:openhands-interaction-cost}
\vspace{-2mm}
\renewcommand{\arraystretch}{1.08}
\setlength{\tabcolsep}{3.0pt}

\begin{tabular}{
L{1.75cm}
C{0.68cm}
C{0.58cm}
C{0.70cm}
C{0.92cm}
}

\toprule[0.8pt]

\cellcolor{interactionheadbg}\textbf{Model}
&
\cellcolor{interactionheadbg}\textbf{Pass@1}
&
\cellcolor{interactionheadbg}\textbf{Steps}
&
\cellcolor{interactionheadbg}\textbf{Tokens}
&
\cellcolor{interactionheadbg}\textbf{Time}
\\

\arrayrulecolor{mainheaderblue}
\midrule
\arrayrulecolor{black}
Qwen 3.6 Plus    & 4.09  & 258 & 12.4M & 1h 12m \\
Kimi K2.6        & 8.36  & 159 & 9.0M  & 2h 17m \\
MiniMax M2.7     & 6.78  & 279 & 24.4M & 56m 12s \\
GPT-5.4          & 7.44  & 36  & 1.4M  & 7m 16s \\
Gemini 3.1 Pro   & 8.10  & 84  & 4.1M  & 12m 52s \\
DeepSeek V4 Pro  & 10.97 & 223 & 9.8M  & 1h 24m \\
GLM 5.1          & 10.23 & 218 & 8.2M  & 1h 52m \\
Claude Opus 4.7  & 18.12 & 102 & 12.1M & 29m 55s \\

\bottomrule[0.8pt]

\end{tabular}

\end{wraptable}
Due to the high complexity of SaaSBench tasks, agents often perform long autonomous multi-turn interactions. As shown in Table~\ref{tab:openhands-interaction-cost} (a), we report steps, total token consumption, and average time to analyze interaction behavior and execution cost in end-to-end SaaS construction. We find that more steps do not guarantee higher system completion. Successful construction depends on the ability of the agent to use environmental feedback, identify root causes of build errors, runtime exceptions, interface responses, and data-state issues, and apply targeted fixes.

Specifically, we observe two typical phenomena. \textbf{The first is insufficient interaction or premature convergence:} GPT-5.4 executes only 36 steps yet achieves a score of 7.44\%, suggesting high single-step generation quality but limited long-horizon debugging and system validation. \textbf{The second is ineffective long-horizon interaction:} MiniMax M2.7 executes 279 steps yet achieves only 6.78\%, indicating that many attempts may still result in repeated debugging, local patching, or inefficient exploration. These results show that long-horizon SaaS development depends more on high-quality reason-act-observe loops than on a larger interaction budget.

\subsection{Error Analysis and Failure Modes}
\label{Error_Analysis}
To examine where the development process breaks down, we analyze 480 capability units from two agents and define five execution-trajectory categories, as detailed in Appendix~\ref{app:failure-mode-taxonomy}. As shown in Figure~\ref{tab:openhands-interaction-cost} (b), the dominant failure mode in SaaSBench is not weakness in a single capability dimension. Instead, most failures occur before agents reach deep business logic. Overall, in 63.5\% of the capability units, the generated stack never runs stably. Another 32.1\% are only superficially accessible but structurally incomplete. Only 3.8\% progress to a stage where incomplete business logic becomes the main bottleneck. This indicates that real enterprise-level SaaS development first tests process discipline, deployment stability, dependency management, schema correctness, and reproducible execution, rather than isolated algorithmic or API skills.
Therefore, the key limitation of current state-of-the-art coding agents is not failure to implement specific advanced business logic. Rather, they lack the ability to stably complete end-to-end engineering setup. This is exactly the challenge that future coding agents must overcome.

\section{Conclusion}
In this paper, we introduce \textbf{SaaSBench}, a comprehensive benchmark for evaluating the ability of coding agents to develop and deploy enterprise-level SaaS systems from scratch. Through extensive experiments, we find that even Claude Opus 4.7, the strongest model in our evaluation, performs poorly in delivering a complete system from scratch. We hope that SaaSBench can provide an important foundation for the future development of coding agents and help move the field toward practical ``Vibe Coding''.

%\newpage
% \section*{References}
\bibliography{main}

@String(AAAI  = {AAAI})

@article{chen2021evaluating,
  author       = {Mark Chen and
                  Jerry Tworek and
                  Heewoo Jun and
                  Qiming Yuan and
                  Henrique Pond{\'{e}} de Oliveira Pinto and
                  Jared Kaplan and
                  Harri Edwards and
                  Yuri Burda and
                  Nicholas Joseph and
                  Greg Brockman and
                  Alex Ray and
                  Raul Puri and
                  Gretchen Krueger and
                  Michael Petrov and
                  Heidy Khlaaf and
                  Girish Sastry and
                  Pamela Mishkin and
                  Brooke Chan and
                  Scott Gray and
                  Nick Ryder and
                  Mikhail Pavlov and
                  Alethea Power and
                  Lukasz Kaiser and
                  Mohammad Bavarian and
                  Clemens Winter and
                  Philippe Tillet and
                  Felipe Petroski Such and
                  Dave Cummings and
                  Matthias Plappert and
                  Fotios Chantzis and
                  Elizabeth Barnes and
                  Ariel Herbert{-}Voss and
                  William Hebgen Guss and
                  Alex Nichol and
                  Alex Paino and
                  Nikolas Tezak and
                  Jie Tang and
                  Igor Babuschkin and
                  Suchir Balaji and
                  Shantanu Jain and
                  William Saunders and
                  Christopher Hesse and
                  Andrew N. Carr and
                  Jan Leike and
                  Joshua Achiam and
                  Vedant Misra and
                  Evan Morikawa and
                  Alec Radford and
                  Matthew Knight and
                  Miles Brundage and
                  Mira Murati and
                  Katie Mayer and
                  Peter Welinder and
                  Bob McGrew and
                  Dario Amodei and
                  Sam McCandlish and
                  Ilya Sutskever and
                  Wojciech Zaremba},
  title        = {Evaluating Large Language Models Trained on Code},
  journal      = {CoRR},
  volume       = {abs/2107.03374},
  year         = {2021},
  url          = {https://arxiv.org/abs/2107.03374},
  eprinttype   = {arXiv},
  eprint       = {2107.03374}
}

@article{gao2025trae,
  author       = {Pengfei Gao and
                  Zhao Tian and
                  Xiangxin Meng and
                  Xinchen Wang and
                  Ruida Hu and
                  Yuanan Xiao and
                  Yizhou Liu and
                  Zhao Zhang and
                  Junjie Chen and
                  Cuiyun Gao and
                  Yun Lin and
                  Yingfei Xiong and
                  Chao Peng and
                  Xia Liu},
  title        = {Trae Agent: An LLM-based Agent for Software Engineering with Test-time
                  Scaling},
  journal      = {CoRR},
  volume       = {abs/2507.23370},
  year         = {2025},
  url          = {https://arxiv.org/abs/2507.23370},
  eprinttype   = {arXiv},
  eprint       = {2507.23370}
}

@article{lin2025se,
  author       = {Jiaye Lin and
                  Yifu Guo and
                  Yuzhen Han and
                  Sen Hu and
                  Ziyi Ni and
                  Licheng Wang and
                  Mingguang Chen and
                  Hongzhang Liu and
                  Ronghao Chen and
                  Yangfan He and
                  Daxin Jiang and
                  Binxing Jiao and
                  Chen Hu and
                  Huacan Wang},
  title        = {SE-Agent: Self-Evolution Trajectory Optimization in Multi-Step Reasoning
                  with LLM-Based Agents},
  journal      = {CoRR},
  volume       = {abs/2508.02085},
  year         = {2025},
  url          = {https://arxiv.org/abs/2508.02085},
  eprinttype   = {arXiv},
  eprint       = {2508.02085}
}

@article{austin2021program,
  author       = {Jacob Austin and
                  Augustus Odena and
                  Maxwell Nye and
                  Maarten Bosma and
                  Henryk Michalewski and
                  David Dohan and
                  Ellen Jiang and
                  Carrie Cai and
                  Michael Terry and
                  Quoc Le and
                  Charles Sutton},
  title        = {Program Synthesis with Large Language Models},
  journal      = {CoRR},
  volume       = {abs/2108.07732},
  year         = {2021},
  url          = {https://arxiv.org/abs/2108.07732},
  eprinttype   = {arXiv},
  eprint       = {2108.07732}
}

@article{hendrycks2021measuring,
  author       = {Dan Hendrycks and
                  Steven Basart and
                  Saurav Kadavath and
                  Mantas Mazeika and
                  Akul Arora and
                  Ethan Guo and
                  Collin Burns and
                  Samir Puranik and
                  Horace He and
                  Dawn Song and
                  Jacob Steinhardt},
  title        = {Measuring Coding Challenge Competence With {APPS}},
  journal      = {CoRR},
  volume       = {abs/2105.09938},
  year         = {2021},
  url          = {https://arxiv.org/abs/2105.09938},
  eprinttype   = {arXiv},
  eprint       = {2105.09938}
}

@inproceedings{liu2023repobench,
  author       = {Tianyang Liu and
                  Canwen Xu and
                  Julian McAuley},
  title        = {RepoBench: Benchmarking Repository-Level Code Auto-Completion Systems},
  booktitle    = {The Twelfth International Conference on Learning Representations},
  year         = {2024},
  url          = {https://openreview.net/forum?id=pPjZIOuQuF}
}

@article{jimenez2023swe,
  author       = {Carlos E. Jimenez and
                  John Yang and
                  Alexander Wettig and
                  Shunyu Yao and
                  Kexin Pei and
                  Ofir Press and
                  Karthik Narasimhan},
  title        = {SWE-bench: Can Language Models Resolve Real-World GitHub Issues?},
  journal      = {CoRR},
  volume       = {abs/2310.06770},
  year         = {2024},
  url          = {https://arxiv.org/abs/2310.06770},
  eprinttype   = {arXiv},
  eprint       = {2310.06770}
}

@inproceedings{liu2025projecteval,
  author       = {Kaiyuan Liu and
                  Youcheng Pan and
                  Yang Xiang and
                  Daojing He and
                  Jing Li and
                  Yexing Du and
                  Tianrun Gao},
  title        = {ProjectEval: {A} Benchmark for Programming Agents Automated Evaluation
                  on Project-Level Code Generation},
  booktitle    = {Findings of the Association for Computational Linguistics},
  pages        = {20205--20221},
  year         = {2025},
  url          = {https://aclanthology.org/2025.findings-acl.1036/}
}

@article{ding2026nl2repobenchlonghorizonrepositorygeneration,
  author={Jingzhe Ding and Shengda Long and Changxin Pu and Huan Zhou and Hongwan Gao and Xiang Gao and Chao He and Yue Hou and Fei Hu and Zhaojian Li and Weiran Shi and Zaiyuan Wang and Daoguang Zan and Chenchen Zhang and Xiaoxu Zhang and Qizhi Chen and Xianfu Cheng and Bo Deng and Qingshui Gu and Kai Hua and Juntao Lin and Pai Liu and Mingchen Li and Xuanguang Pan and Zifan Peng and Yujia Qin and Yong Shan and Zhewen Tan and Weihao Xie and Zihan Wang and Yishuo Yuan and Jiayu Zhang and Enduo Zhao and Yunfei Zhao and He Zhu and Liya Zhu and Chenyang Zou and Ming Ding and Jianpeng Jiao and Jiaheng Liu and Minghao Liu and Qian Liu and Chongyang Tao and Jian Yang and Tong Yang and Zhaoxiang Zhang and Xinjie Chen and Wenhao Huang and Ge Zhang},
  title        = {NL2Repo-Bench: Towards Long-Horizon Repository Generation Evaluation
                  of Coding Agents},
  journal      = {CoRR},
  volume       = {abs/2512.12730},
  year         = {2026},
  url          = {https://arxiv.org/abs/2512.12730},
  eprinttype   = {arXiv},
  eprint       = {2512.12730}
}

@article{peng2026repogenesis,
  author       = {Zhiyuan Peng and
                  Xin Yin and
                  Pu Zhao and
                  Fangkai Yang and
                  Lu Wang and
                  Ran Jia and
                  Xu Chen and
                  Qingwei Lin and
                  Saravan Rajmohan and
                  Dongmei Zhang},
  title        = {RepoGenesis: Benchmarking End-to-End Microservice Generation from
                  Readme to Repository},
  journal      = {CoRR},
  volume       = {abs/2601.13943},
  year         = {2026},
  url          = {https://arxiv.org/abs/2601.13943},
  eprinttype   = {arXiv},
  eprint       = {2601.13943}
}

@article{fu2025automatically,
  author       = {Lingyue Fu and
                  Bolun Zhang and
                  Hao Guan and
                  Yaoming Zhu and
                  Lin Qiu and
                  Weiwen Liu and
                  Xuezhi Cao and
                  Xunliang Cai and
                  Weinan Zhang and
                  Yong Yu},
  title        = {Automatically Benchmarking {LLM} Code Agents through Agent-Driven
                  Annotation and Evaluation},
  journal      = {CoRR},
  volume       = {abs/2510.24358},
  year         = {2025},
  url          = {https://arxiv.org/abs/2510.24358},
  eprinttype   = {arXiv},
  eprint       = {2510.24358}
}

@article{lu2026projdevbench,
  author       = {Pengrui Lu and
                  Shiqi Zhang and
                  Yunzhong Hou and
                  Lyumanshan Ye and
                  Chaoyi Huang and
                  Zixi Chen and
                  Ji Zeng and
                  Hantao Jiang and
                  Pengfei Liu and
                  Yiwei Wang and
                  Ming{-}Hsuan Yang},
  title        = {ProjDevBench: Benchmarking {AI} Coding Agents on End-to-End Project
                  Development},
  journal      = {CoRR},
  volume       = {abs/2602.01655},
  year         = {2026},
  url          = {https://arxiv.org/abs/2602.01655},
  eprinttype   = {arXiv},
  eprint       = {2602.01655}
}

@article{li2022alphacode,
  author = {Li, Yujia and Choi, David and Chung, Junyoung and Kushman, Nate and Schrittwieser, Julian and Leblond, R{\'e}mi and Eccles, Tom and Keeling, James and Gimeno, Felix and Lago, Agustin Dal and Hubert, Thomas and Choy, Peter and de Masson d’Autume, Cyprien and Babuschkin, Igor and Chen, Xinyun and Huang, Po-Sen and Welbl, Johannes and Gowal, Sven and Cherepanov, Alexey and Molloy, James and Mankowitz, Daniel J. and Sutherland Robson, Esme and Kohli, Pushmeet and de Freitas, Nando and Kavukcuoglu, Koray and Vinyals, Oriol},
  title = {Competition-level code generation with AlphaCode},
  journal = {Science},
  volume = {378},
  number = {6624},
  pages = {1092--1097},
  year = {2022},
  doi = {10.1126/science.abq1158},
  url = {https://www.science.org/doi/abs/10.1126/science.abq1158}
}

@article{dong2025survey,
  author       = {Yihong Dong and
                  Xue Jiang and
                  Jiaru Qian and
                  Tian Wang and
                  Kechi Zhang and
                  Zhi Jin and
                  Ge Li},
  title        = {A Survey on Code Generation with LLM-based Agents},
  journal      = {CoRR},
  volume       = {abs/2508.00083},
  year         = {2025},
  url          = {https://arxiv.org/abs/2508.00083},
  eprinttype   = {arXiv},
  eprint       = {2508.00083}
}

@article{ge2025survey,
  author       = {Yuyao Ge and
                  Lingrui Mei and
                  Zenghao Duan and
                  Tianhao Li and
                  Yujia Zheng and
                  Yiwei Wang and
                  Lexin Wang and
                  Jiayu Yao and
                  Tianyu Liu and
                  Yujun Cai and
                  Baolong Bi and
                  Fangda Guo and
                  Jiafeng Guo and
                  Shenghua Liu and
                  Xueqi Cheng},
  title        = {A Survey of Vibe Coding with Large Language Models},
  journal      = {CoRR},
  volume       = {abs/2510.12399},
  year         = {2025},
  url          = {https://arxiv.org/abs/2510.12399},
  eprinttype   = {arXiv},
  eprint       = {2510.12399}
}

@article{sapkota2025vibe,
  author       = {Ranjan Sapkota and
                  Konstantinos I. Roumeliotis and
                  Manoj Karkee},
  title        = {Vibe Coding vs. Agentic Coding: Fundamentals and Practical Implications
                  of Agentic {AI}},
  journal      = {CoRR},
  volume       = {abs/2505.19443},
  year         = {2025},
  url          = {https://arxiv.org/abs/2505.19443},
  eprinttype   = {arXiv},
  eprint       = {2505.19443}
}

@article{sarkar2025vibe,
  title={Vibe coding: programming through conversation with artificial intelligence},
  author={Sarkar, Advait and Drosos, Ian},
  journal={arXiv preprint arXiv:2506.23253},
  year={2025},
  url={https://arxiv.org/abs/2506.23253}
}

@inproceedings{wang2024openhands,
    author={Xingyao Wang and Boxuan Li and Yufan Song and Frank F. Xu and Xiangru Tang and Mingchen Zhuge and Jiayi Pan and Yueqi Song and Bowen Li and Jaskirat Singh and Hoang H. Tran and Fuqiang Li and Ren Ma and Mingzhang Zheng and Bill Qian and Yanjun Shao and Niklas Muennighoff and Yizhe Zhang and Binyuan Hui and Junyang Lin and Robert Brennan and Hao Peng and Heng Ji and Graham Neubig},
  title        = {OpenHands: An Open Platform for {AI} Software Developers as Generalist
                  Agents},
  booktitle    = {The Thirteenth International Conference on Learning Representations},
  year         = {2025},
  url          = {https://openreview.net/forum?id=OJd3ayDDoF}
}

@article{deng2025swe,
  author={Xiang Deng and Jeff Da and Edwin Pan and Yannis Yiming He and Charles Ide and Kanak Garg and Niklas Lauffer and Andrew Park and Nitin Pasari and Chetan Rane and Karmini Sampath and Maya Krishnan and Srivatsa Kundurthy and Sean Hendryx and Zifan Wang and Vijay Bharadwaj and Jeff Holm and Raja Aluri and Chen Bo Calvin Zhang and Noah Jacobson and Bing Liu and Brad Kenstler},
  title        = {SWE-Bench Pro: Can {AI} Agents Solve Long-Horizon Software Engineering
                  Tasks?},
  journal      = {CoRR},
  volume       = {abs/2509.16941},
  year         = {2025},
  url          = {https://arxiv.org/abs/2509.16941},
  eprinttype   = {arXiv},
  eprint       = {2509.16941}
}

@misc{cursor2024cursor,
  author       = {{Cursor AI}},
  title        = {Cursor: The AI Code Editor},
  year         = {2024},
  url          = {https://www.cursor.com},
  note         = {Accessed: 2026-05-03}
}

@misc{anthropic2025opus4sonnet4,
  author       = {{Anthropic}},
  title        = {System Card: Claude Opus 4 \& Claude Sonnet 4},
  year         = {2025},
  url = {https://www-cdn.anthropic.com/4263b940cabb546aa0e3283f35b686f4f3b2ff47.pdf},
  note         = {Accessed: 2026-05-03}
}

@misc{anthropic2025sonnet45,
  author       = {{Anthropic}},
  title        = {System Card: Claude Sonnet 4.5},
  year         = {2025},
  url = {https://assets.anthropic.com/m/12f214efcc2f457a/original/Claude-Sonnet-4-5-System-Card.pdf},
  note         = {Accessed: 2026-05-03}
}

@article{liu2025deepseek,
  title={Deepseek-v3.2: Pushing the frontier of open large language models},
  author={Liu, Aixin and Mei, Aoxue and Lin, Bangcai and Xue, Bing and Wang, Bingxuan and Xu, Bingzheng and Wu, Bochao and Zhang, Bowei and Lin, Chaofan and Dong, Chen and others},
  journal={arXiv preprint arXiv:2512.02556},
  year={2025},
  url={https://arxiv.org/abs/2512.02556}
}

@inproceedings{huang2025opencoder,
  author       = {Siming Huang and
                  Tianhao Cheng and
                  Jason Klein Liu and
                  Weidi Xu and
                  Jiaran Hao and
                  Liuyihan Song and
                  Yang Xu and
                  Jian Yang and
                  Jiaheng Liu and
                  Chenchen Zhang and
                  Linzheng Chai and
                  Ruifeng Yuan and
                  Xianzhen Luo and
                  Qiufeng Wang and
                  YuanTao Fan and
                  Qingfu Zhu and
                  Zhaoxiang Zhang and
                  Yang Gao and
                  Jie Fu and
                  Qian Liu and
                  Houyi Li and
                  Ge Zhang and
                  Yuan Qi and
                  Yinghui Xu and
                  Wei Chu and
                  Zili Wang},
  title        = {OpenCoder: The Open Cookbook for Top-Tier Code Large Language Models},
  booktitle    = {Proceedings of the 63rd Annual Meeting of the Association for Computational
                  Linguistics (Volume 1: Long Papers)},
  pages        = {33167--33193},
  year         = {2025},
  url          = {https://aclanthology.org/2025.acl-long.1591/}
}

@inproceedings{li2025prompting,
    title = "Prompting Large Language Models to Tackle the Full Software Development Lifecycle: A Case Study",
    author = "Li, Bowen  and
      Wu, Wenhan  and
      Tang, Ziwei  and
      Shi, Lin  and
      Yang, John  and
      Li, Jinyang  and
      Yao, Shunyu  and
      Qian, Chen  and
      Hui, Binyuan  and
      Zhang, Qicheng  and
      Yu, Zhiyin  and
      Du, He  and
      Yang, Ping  and
      Lin, Dahua  and
      Peng, Chao  and
      Chen, Kai",
    booktitle = "Proceedings of the 31st International Conference on Computational Linguistics",
    year = "2025",
    url = "https://aclanthology.org/2025.coling-main.502/",
    pages = "7511--7531",
}

@misc{github2021copilot,
  author       = {{GitHub}},
  title        = {GitHub Copilot: Your AI Pair Programmer},
  year         = {2021},
  url = {https://copilot.github.com/},
  note         = {Accessed: 2026-05-03}
}

@misc{openai2025codexcli,
  author       = {{OpenAI}},
  title        = {Codex CLI},
  year         = {2025},
  url = {https://github.com/openai/codex},
  note         = {Accessed: 2026-05-03}
}

@misc{anthropic2025claudecodeagentic,
  author       = {{Anthropic}},
  title        = {Claude code: Ai-powered coding assistant},
  year         = {2024},
  url = {https://www.claude.com/product/claude-code},
  note         = {Accessed: 2026-05-03}
}

@inproceedings{ni2026gittaskbench,
  author       = {Ziyi Ni and
                  Huacan Wang and
                  Shuo Zhang and
                  Shuo Lu and
                  Ziyang He and
                  Wang You and
                  Zhenheng Tang and
                  Sen Hu and
                  Bo Li and
                  Chen Hu and
                  Binxing Jiao and
                  Daxin Jiang and
                  Yuntao Du and
                  Pin Lyu},
  title        = {GitTaskBench: {A} Benchmark for Code Agents Solving Real-World Tasks
                  Through Code Repository Leveraging},
  booktitle    = {Proceedings of the AAAI Conference on Artificial Intelligence},
  pages        = {32564--32572},
  year         = {2026},
  url          = {https://doi.org/10.1609/aaai.v40i38.40533}
}

@article{he2025swe,
  title={Swe-perf: Can language models optimize code performance on real-world repositories?},
  author={He, Xinyi and Liu, Qian and Du, Mingzhe and Yan, Lin and Fan, Zhijie and Huang, Yiming and Yuan, Zejian and Ma, Zejun},
  journal={arXiv preprint arXiv:2507.12415},
  year={2025},
  url={https://arxiv.org/abs/2507.12415}
}

@article{yang2024sweagent,
  title={Swe-agent: Agent-computer interfaces enable automated software engineering},
  author={Yang, John and Jimenez, Carlos E and Wettig, Alexander and Lieret, Kilian and Yao, Shunyu and Narasimhan, Karthik and Press, Ofir},
  journal={Advances in Neural Information Processing Systems},
  volume={37},
  pages={50528--50652},
  year={2024},
  url={https://proceedings.neurips.cc/paper_files/paper/2024/hash/5a7c947568c1b1328ccc5230172e1e7c-Abstract-Conference.html}
}

@inproceedings{hong2024metagpt,
  author       = {Sirui Hong and
                  Mingchen Zhuge and
                  Jonathan Chen and
                  Xiawu Zheng and
                  Yuheng Cheng and
                  Jinlin Wang and
                  Ceyao Zhang and
                  Zili Wang and
                  Steven Ka Shing Yau and
                  Zijuan Lin and
                  Liyang Zhou and
                  Chenyu Ran and
                  Lingfeng Xiao and
                  Chenglin Wu and
                  J{\"{u}}rgen Schmidhuber},
  title        = {MetaGPT: Meta Programming for {A} Multi-Agent Collaborative Framework},
  booktitle    = {The Twelfth International Conference on Learning Representations},
  year         = {2024},
  url          = {https://openreview.net/forum?id=VtmBAGCN7o}
}

@inproceedings{pmlr-v267-miserendino25a,
  title = 	 {{SWE}-Lancer: Can Frontier {LLM}s Earn \$1 Million from Real-World Freelance Software Engineering?},
  author =       {Miserendino, Samuel and Wang, Michele and Patwardhan, Tejal and Heidecke, Johannes},
  booktitle = 	 {Proceedings of the 42nd International Conference on Machine Learning},
  pages = 	 {44412--44450},
  year = 	 {2025},
  volume = 	 {267},
  publisher =    {PMLR},
  url = 	 {https://proceedings.mlr.press/v267/miserendino25a.html}
}

@article{xu2025swecompassunifiedevaluationagentic,
  author       = {Jingxuan Xu and
                  Ken Deng and
                  Weihao Li and
                  Songwei Yu and
                  Huaixi Tang and
                  Haoyang Huang and
                  Zhiyi Lai and
                  Zizheng Zhan and
                  Yanan Wu and
                  Chenchen Zhang and
                  Kepeng Lei and
                  Yifan Yao and
                  Xinping Lei and
                  Wenqiang Zhu and
                  Zong{-}Xian Feng and
                  Han Li and
                  Junqi Xiong and
                  Dailin Li and
                  Zuchen Gao and
                  Kun Wu and
                  Wen Xiang and
                  Ziqi Zhan and
                  Yuanxing Zhang and
                  Wuxuan Gong and
                  Ziyuan Gao and
                  Guanxiang Wang and
                  Yirong Xue and
                  Mengtong Li and
                  Mengfei Xie and
                  Xiaojiang Zhang and
                  Jinghui Wang and
                  Wenhao Zhuang and
                  Zheng Lin and
                  Huiming Wang and
                  Zhaoxiang Zhang and
                  Yuqun Zhang and
                  Haotian Zhang and
                  Bin Chen and
                  Jiaheng Liu},
  title        = {SWE-Compass: Towards Unified Evaluation of Agentic Coding Abilities
                  for Large Language Models},
  journal      = {CoRR},
  volume       = {abs/2511.05459},
  year         = {2025},
  url          = {https://arxiv.org/abs/2511.05459},
  eprinttype   = {arXiv},
  eprint       = {2511.05459}
}

@inproceedings{liu2025m2rc,
  author       = {Jiaheng Liu and
                  Ken Deng and
                  Congnan Liu and
                  Jian Yang and
                  Shukai Liu and
                  He Zhu and
                  Peng Zhao and
                  Linzheng Chai and
                  Yanan Wu and
                  Ke Jin and
                  Ge Zhang and
                  Zekun Moore Wang and
                  Guoan Zhang and
                  Yingshui Tan and
                  Bangyu Xiang and
                  Zhaoxiang Zhang and
                  Wenbo Su and
                  Bo Zheng},
  title        = {{M2RC-EVAL:} Massively Multilingual Repository-level Code Completion
                  Evaluation},
  booktitle    = {Proceedings of the 63rd Annual Meeting of the Association for Computational
                  Linguistics (Volume 1: Long Papers)},
  pages        = {15661--15684},
  year         = {2025},
  url          = {https://aclanthology.org/2025.acl-long.763/}
}

@article{CodeContests+,
  author       = {Zihan Wang and
                  Siyao Liu and
                  Yang Sun and
                  Hongyan Li and
                  Kai Shen},
  title        = {CodeContests+: High-Quality Test Case Generation for Competitive Programming},
  journal      = {CoRR},
  volume       = {abs/2506.05817},
  year         = {2025},
  url = {https://arxiv.org/abs/2506.05817},
  eprinttype   = {arXiv},
  eprint       = {2506.05817}
}

@article{EvoCodeBench,
  author       = {Jia Li and
                  Ge Li and
                  Xuanming Zhang and
                  Yihong Dong and
                  Zhi Jin},
  title        = {EvoCodeBench: An Evolving Code Generation Benchmark Aligned with Real-World
                  Code Repositories},
  journal      = {CoRR},
  volume       = {abs/2404.00599},
  year         = {2024},
  url          = {https://arxiv.org/abs/2404.00599},
  eprinttype   = {arXiv},
  eprint       = {2404.00599},
}

@inproceedings{lai2023ds,
  title={DS-1000: A natural and reliable benchmark for data science code generation},
  author={Lai, Yuhang and Li, Chengxi and Wang, Yiming and Zhang, Tianyi and Zhong, Ruiqi and Zettlemoyer, Luke and Yih, Wen-tau and Fried, Daniel and Wang, Sida and Yu, Tao},
  booktitle={International Conference on Machine Learning},
  pages={18319--18345},
  year={2023},
  organization={PMLR},
  url={https://proceedings.mlr.press/v202/lai23b.html}
}

@article{WebGen-Bench,
  author       = {Zimu Lu and
                  Yunqiao Yang and
                  Houxing Ren and
                  Haotian Hou and
                  Han Xiao and
                  Ke Wang and
                  Weikang Shi and
                  Aojun Zhou and
                  Mingjie Zhan and
                  Hongsheng Li},
  title        = {WebGen-Bench: Evaluating LLMs on Generating Interactive and Functional
                  Websites from Scratch},
  journal      = {CoRR},
  volume       = {abs/2505.03733},
  year         = {2025},
  url          = {https://arxiv.org/abs/2505.03733},
  eprinttype   = {arXiv},
  eprint       = {2505.03733}
}

@article{jiang2026survey,
  title={A survey on large language models for code generation},
  author={Jiang, Juyong and Wang, Fan and Shen, Jiasi and Kim, Sungju and Kim, Sunghun},
  journal={ACM Transactions on Software Engineering and Methodology},
  volume={35},
  number={2},
  pages={1--72},
  year={2026},
  publisher={ACM New York, NY},
  url={https://dl.acm.org/doi/10.1145/3747588}
}

@inproceedings{BigCodeBench,
  author={Terry Yue Zhuo and Minh Chien Vu and Jenny Chim and Han Hu and Wenhao Yu and Ratnadira Widyasari and Imam Nur Bani Yusuf and Haolan Zhan and Junda He and Indraneil Paul and Simon Brunner and Chen Gong and Thong Hoang and Armel Randy Zebaze and Xiaoheng Hong and Wen-Ding Li and Jean Kaddour and Ming Xu and Zhihan Zhang and Prateek Yadav and Naman Jain and Alex Gu and Zhoujun Cheng and Jiawei Liu and Qian Liu and Zijian Wang and Binyuan Hui and Niklas Muennighoff and David Lo and Daniel Fried and Xiaoning Du and Harm de Vries and Leandro Von Werra},
  title        = {BigCodeBench: Benchmarking Code Generation with Diverse Function Calls
                  and Complex Instructions},
  booktitle    = {The Thirteenth International Conference on Learning Representations},
  year         = {2025},
  url          = {https://openreview.net/forum?id=YrycTjllL0},
}

@misc{openai2026gpt54,
  author       = {{OpenAI}},
  title        = {{GPT-5.4 Thinking System Card}},
  year         = {2026},
  url = {https://deploymentsafety.openai.com/gpt-5-4-thinking/gpt-5-4-thinking.pdf},
  note         = {Accessed: 2026-05-03}
}

@misc{google2026gemini31propreview,
  author       = {{The Gemini Team}},
  title        = {{Gemini 3.1 Pro: }A smarter model for your most complex tasks},
  year         = {2026},
  url = {https://blog.google/innovation-and-ai/models-and-research/gemini-models/gemini-3-1-pro/},
  note         = {Accessed: 2026-05-03}
}

@misc{anthropic2026claudeopus47,
  author       = {{Anthropic}},
  title        = {{Introducing Claude Opus 4.7}},
  year         = {2026},
  url = {https://www.anthropic.com/news/claude-opus-4-7},
  note         = {Accessed: 2026-05-03}
}

@misc{moonshotai2026kimik26,
  author       = {{Moonshot AI}},
  title        = {{Kimi K2.6}: Advancing Open-Source Coding},
  year         = {2026},
  url = {https://www.kimi.com/blog/kimi-k2-6},
  note         = {Accessed: 2026-05-03}
}

@misc{qwen2026qwen36plus,
  author       = {{Qwen Team}},
  title        = {{Qwen3.6-Plus}: Towards Real World Agents},
  year         = {2026},
  url = {https://qwen.ai/blog?id=qwen3.6},
  note         = {Accessed: 2026-05-03}
}

@misc{zai2026glm51,
  author       = {{Z.AI}},
  title        = {GLM-5.1: Towards Long-Horizon Tasks},
  year         = {2026},
  url = {https://z.ai/blog/glm-5.1},
  note         = {Accessed: 2026-05-03}
}

@misc{minimax2026m27,
  author       = {{MiniMax}},
  title        = {{MiniMax M2.7}: Early Echoes of Self-Evolution},
  year         = {2026},
  url = {https://www.minimax.io/news/minimax-m27-en},
  note         = {Accessed: 2026-05-03}
}

@misc{qwen3.6-27b,
    title = {{Qwen3.6-27B}: Flagship-Level Coding in a {27B} Dense Model},
    author = {{Qwen Team}},
    year = {2026},
    url = {https://qwen.ai/blog?id=qwen3.6-27b},
    note         = {Accessed: 2026-05-03}
}

@techreport{deepseekai2026deepseekv4,
  title        = {DeepSeek-V4: Towards Highly Efficient Million-Token Context Intelligence},
  author       = {{DeepSeek-AI}},
  institution  = {DeepSeek-AI},
  year         = {2026},
  type         = {Technical report},
  url          = {https://huggingface.co/deepseek-ai/DeepSeek-V4-Pro/blob/main/DeepSeek_V4.pdf},
  note         = {Accessed: 2026-05-03}
}

%%%%%%%%%%%%%%%%%%%%%%%%%%%%%%%%%%%%%%%%%%%%%%%%%%%%%%%%%%%%

\appendix
\appendix

\section{Benchmark Details and Statistics}
\definecolor{saashead}{HTML}{1F3A93}
\definecolor{saasalt}{HTML}{F4F6FA}
\definecolor{domlabel}{HTML}{1F3A93}

\begin{table*}[th]
\centering
\scriptsize
\renewcommand{\arraystretch}{1.18}
\setlength{\tabcolsep}{4pt}
\caption{\textbf{SaaSBench category definitions and engineering scope (Part I).}
Rows 1--15 summarise the functional scope and distinctive engineering
challenge of each category.}
\label{tab:saasbench-category-defs-a}
\rowcolors{2}{white}{saasalt}
\resizebox{\textwidth}{!}{%
\begin{tabular}{@{}c l p{3.2cm} p{5.0cm} p{5.2cm}@{}}
\toprule
\textbf{\#} & \textbf{Dom.} & \textbf{Category} &
\textbf{Definition} & \textbf{Core Engineering Challenge} \\
\midrule
1 & CG & Email \& Newsletter &
SaaS that composes, segments, and delivers bulk transactional and
marketing email, and tracks open/click/bounce funnels. &
High-throughput SMTP pipeline with DKIM/SPF/DMARC signing, bounce/
complaint handling, and per-tenant IP-warming and reputation. \\
2 & CG & Customer Relationship Management &
System of record for accounts, contacts, opportunities and deal
pipelines, with activity logging and forecasting. &
Multi-object relational graph (account/contact/opportunity/activity)
with custom fields, role-based views, and pipeline state machines. \\
3 & CG & Help Desk \& Ticketing &
Omni-channel inbox (email, chat, voice, social) that unifies customer
requests as tickets with SLAs, macros, and agent workflows. &
Channel unification, SLA/priority timers, skill-based routing, and
agent collision detection across concurrent conversations. \\
4 & CG & Form Builder \& Survey &
No-code builder for web forms, surveys, and quizzes with conditional
logic and response analytics. &
Conditional-logic branching engine, schema versioning, and response
ingestion with partial-submit and anti-fraud controls. \\
5 & CG & Community Forum &
Threaded public/private discussion platform with trust levels,
moderation tooling, and topic-level SEO. &
Trust-level reputation model, spam/abuse moderation queue, and
topic ranking with SEO-grade public rendering. \\
6 & CG & Real-time Communication &
Team messaging with channels, threads, direct messages, presence,
and file sharing, over persistent WebSocket. &
Persistent WebSocket fan-out, channel/thread model, presence and
typing indicators, and federation across workspaces. \\
7 & CG & Video Conferencing &
Real-time audio/video meetings with screen share, recording, and
large-room broadcasting. &
WebRTC SFU media routing with simulcast, bandwidth adaptation,
and server-side recording / transcoding. \\
8 & PC & Project Management \& Issue Tracking &
Tool for planning, tracking, and reporting work items across sprints,
epics, and cross-functional teams. &
Multi-view (board / list / timeline / Gantt) synchronisation, sprint
engine, and dependency graph with cycle detection. \\
9 & PC & Knowledge Base \& Wiki &
Hierarchical collaborative document system with rich-text editing,
versioning, search, and permissioned sharing. &
Hierarchical page tree, full-text search over Markdown/AST, CRDT-
or OT-style concurrent editing with versioning. \\
10 & PC & Time Tracking &
Tool that records time spent per task/project, generates timesheets,
and exports billable hours. &
Live timer engine with project switching, timesheet aggregation, and
offline / cross-device reconciliation. \\
11 & PC & Gamified Productivity &
Habit/to-do app that models tasks as RPG quests with rewards,
streaks, and social challenges. &
Deterministic RPG state machine (HP/XP/streaks), economy balance,
and asynchronous social quest consensus. \\
12 & PC & Scheduling \& Booking &
Self-service scheduling links that expose a user's availability and
let external parties book within configurable rules. &
Cross-calendar availability algorithm, timezone-aware conflict
detection, and buffer / round-robin booking policies. \\
13 & PC & Learning Management System &
Platform that authors courses, delivers content, runs assessments,
and reports learner outcomes. &
Course/module engine, quiz grading with item analysis, and SCORM
/ xAPI content interoperability. \\
14 & CF & E-commerce Platform &
Multi-channel storefront with product catalogue, cart, checkout,
payment and order management. &
Cart and order state machine, payment-gateway orchestration, and
inventory consistency under concurrent checkout. \\
15 & CF & Billing \& Subscription &
Engine that meters usage, prices subscription plans, and issues
recurring invoices with dunning and renewals. &
Usage metering at scale, proration and mid-cycle plan changes, and
idempotent invoicing with revenue-recognition safety. \\
\bottomrule
\end{tabular}
}
\end{table*}

\definecolor{saashead}{HTML}{1F3A93}
\definecolor{saasalt}{HTML}{F4F6FA}
\definecolor{domlabel}{HTML}{1F3A93}
\begin{table*}[th]
\centering
\scriptsize
\renewcommand{\arraystretch}{1.18}
\setlength{\tabcolsep}{4pt}
\caption{\textbf{SaaSBench category definitions and engineering scope (Part II).}
Continuation of Table~\ref{tab:saasbench-category-defs-a}.}
\label{tab:saasbench-category-defs-b}
\rowcolors{2}{white}{saasalt}
\resizebox{\textwidth}{!}{%
\begin{tabular}{@{}c l p{3.2cm} p{5.0cm} p{5.2cm}@{}}
\toprule
\textbf{\#} & \textbf{Dom.} & \textbf{Category} &
\textbf{Definition} & \textbf{Core Engineering Challenge} \\
\midrule
16 & CF & Accounting \& Invoicing &
SMB bookkeeping with chart of accounts, invoices, expenses, tax and
financial reports. &
Double-entry ledger invariants, multi-currency conversion and tax
computation, and audit-grade report generation. \\
17 & CF & Inventory \& Warehouse &
System that tracks stock levels, locations, movements, and BOMs
across warehouses. &
Stock-movement ledger, multi-location transfer and BOM hierarchy,
and barcode / label pipeline with scanner integration. \\
18 & DCI & Headless CMS &
API-first content platform where editors model content types and
consume them via REST/GraphQL from any front end. &
Schema-to-API generation, media processing pipeline, draft/publish
workflow, and webhook-driven content delivery. \\
19 & DCI & Business Intelligence &
Tool for exploring data, authoring dashboards, and sharing
interactive visualisations across an organisation. &
SQL authoring / semantic layer, SQL-$\rightarrow$-chart rendering,
and governed multi-source connector fleet. \\
20 & DCI & Data Catalog \& Lineage &
Central registry of datasets, schemas, owners, usage and column-
level lineage across the data stack. &
Metadata ingestion from heterogeneous sources, lineage-graph
construction, and impact-analysis queries. \\
21 & DCI & Feature Flag \& Experimentation &
SDK + console for toggling features by segment and running A/B and
feature-rollout experiments. &
Low-latency flag evaluation at the edge, targeting-rule engine, and
statistical rigour of A/B analysis. \\
22 & DCI & Web \& Product Analytics &
Lightweight tracker that ingests page/event data and reports traffic,
conversion and engagement metrics. &
High-cardinality event ingest, real-time aggregation, and privacy-
preserving (cookieless) attribution. \\
23 & SI & Monitoring \& Security Ops &
Dashboards, alerts and time-series exploration over metrics, logs,
traces and synthetic probes, often doubling as the SOC/SecOps
visualisation layer. &
Probe scheduling, time-series storage and dashboarding, alert-rule
evaluation with notification routing, and log/event correlation for
security operations. \\
24 & SI & Identity \& Access Management &
Centralised identity server that federates OIDC/SAML/LDAP, enforces
MFA, and manages users, roles and realms. &
Protocol coverage (OIDC/SAML/LDAP/SCIM), MFA and session model,
and tenant/realm isolation with delegated administration. \\
25 & SI & Password \& Secrets Management &
Encrypted vault for passwords and secrets with cross-device sync
and organisation-level sharing. &
Zero-knowledge end-to-end encryption, client-side vault model, and
secure cross-device sync / recovery. \\
26 & SI & File Storage \& Sync &
Self-hosted cloud drive with chunked upload, multi-device sync,
sharing links, and collaboration apps. &
Chunked / resumable upload, multi-device delta sync, and external
sharing with link-scoped access control. \\
27 & DW & E-Signature \& Contract Management &
Platform to prepare, route, sign and archive documents with legal
audit trail. &
Signing-order orchestration, tamper-evident hash chain, and audit-
log evidence compliant with eIDAS / ESIGN. \\
28 & DW & Low-Code / No-Code &
Visual builder that assembles internal business apps on top of
existing databases and APIs. &
Visual page/component renderer, dynamic schema binding, and multi-
source (DB/REST/GraphQL) data orchestration. \\
29 & DW & Electronic Health Records &
Clinical record system for patient charts, encounters, prescribing,
and billing. &
FHIR / HL7 interoperability, ICD / CPT / SNOMED coding, and
e-prescribing with clinical-decision support. \\
30 & DW & Workflow Automation &
Visual DAG engine that chains SaaS APIs into automations with
triggers, conditions and error handling. &
Visual DAG orchestration, connector ecosystem, and durable
execution with retries / idempotency. \\
\bottomrule
\end{tabular}
}
\end{table*}

\definecolor{saashead}{HTML}{1F3A93}
\definecolor{saasalt}{HTML}{F4F6FA}
\definecolor{domlabel}{HTML}{1F3A93}
\begin{table*}[th]
\centering
\scriptsize
\renewcommand{\arraystretch}{1.18}
\setlength{\tabcolsep}{4pt}
\caption{\textbf{Market grounding of the SaaSBench categories (Part I).}
Rows 1--15 list the commercial segment, representative analyst
rankings, and flagship proprietary products of each category.}
\label{tab:saasbench-category-defs-c}
\rowcolors{2}{white}{saasalt}
\resizebox{\textwidth}{!}{%
\begin{tabular}{@{}c l p{3.2cm} p{5.8cm} p{4.6cm}@{}}
\toprule
\textbf{\#} & \textbf{Dom.} & \textbf{Category} &
\textbf{Market Segment \& Rankings} &
\textbf{Representative Products} \\
\midrule
1 & CG & Email \& Newsletter &
\textbf{Email Marketing \& Marketing Automation.}
Gartner MQ for B2B Marketing Automation; Forrester Wave for
Email Marketing; G2 Grid for Email Marketing. &
Mailchimp (Intuit); HubSpot Marketing Hub. \\
2 & CG & Customer Relationship Management &
\textbf{CRM / Sales Force Automation.}
Gartner MQ for Sales Force Automation; Forrester Wave for CRM
Suites; IDC\,MS for CRM. &
Salesforce Sales Cloud; Microsoft Dynamics~365 Sales. \\
3 & CG & Help Desk \& Ticketing &
\textbf{Customer Service \& Support.}
Gartner MQ for the CRM Customer Engagement Center; Forrester Wave
for Customer Service Solutions. &
Zendesk; Freshdesk (Freshworks). \\
4 & CG & Form Builder \& Survey &
\textbf{Online Survey \& Form Building.}
G2 Grid for Survey / Online Form Builder; Forrester mentions in
Experience-Management landscape. &
Typeform; SurveyMonkey (Momentive). \\
5 & CG & Community Forum &
\textbf{Online Community Management.}
G2 Grid for Online Community Management; Forrester
Wave for Community Platforms. &
Discourse (commercial cloud); Higher Logic Vanilla. \\
6 & CG & Real-time Communication &
\textbf{Team Collaboration \& Messaging.}
Gartner MQ for Unified Communications as a Service (UCaaS);
G2 Grid for Business Instant Messaging. &
Slack (Salesforce); Microsoft Teams. \\
7 & CG & Video Conferencing &
\textbf{Meeting Solutions / UCaaS.}
Gartner MQ for Meeting Solutions; IDC\,MS for Worldwide UCaaS. &
Zoom Meetings; Cisco Webex. \\
8 & PC & Project Management \& Issue Tracking &
\textbf{Project \& Portfolio Mgmt. / Agile Work Mgmt.}
Gartner MQ for Adaptive Project Mgmt. \& Reporting; Forrester Wave
for Enterprise Agile Planning Tools. &
Atlassian Jira; monday.com Work~OS. \\
9 & PC & Knowledge Base \& Wiki &
\textbf{Knowledge Mgmt. / Collaborative Docs.}
Gartner MQ for Insight Engines (adjacent); G2 Grid for Knowledge
Mgmt. and for Note-Taking Software. &
Notion; Confluence (Atlassian). \\
10 & PC & Time Tracking &
\textbf{Time Tracking \& Professional Services Automation.}
G2 Grid for Time Tracking; Gartner MQ for PSA (adjacent). &
Toggl Track; Harvest. \\
11 & PC & Gamified Productivity &
\textbf{Habit Tracking / Personal Productivity.}
G2 Grid for Task Mgmt. (adjacent); niche leader in the
Quantified-Self / habit-app landscape. &
Habitica (SaaS tier); Todoist (``karma'' system as closest
commercial analog). \\
12 & PC & Scheduling \& Booking &
\textbf{Online Appointment Scheduling.}
G2 Grid for Online Appointment Scheduling; Gartner mention in
Digital Commerce Experience. &
Calendly; Microsoft Bookings. \\
13 & PC & Learning Management System &
\textbf{Learning Management / Corporate LMS.}
Gartner MQ for Higher-Ed \& Corporate LMS; Forrester Wave for
Learning Platforms. &
Canvas (Instructure); Blackboard Learn (Anthology). \\
14 & CF & E-commerce Platform &
\textbf{Digital Commerce.}
Gartner MQ for Digital Commerce; Forrester Wave for
B2B~/~B2C Commerce Solutions. &
Shopify Plus; Adobe Commerce (Magento). \\
15 & CF & Billing \& Subscription &
\textbf{Recurring Billing \& Subscription Management.}
Gartner MQ for Recurring Billing; IDC\,MS for SaaS/Subscription
Billing. &
Stripe Billing; Zuora Billing. \\
\bottomrule
\end{tabular}
}
\end{table*}

\definecolor{saashead}{HTML}{1F3A93}
\definecolor{saasalt}{HTML}{F4F6FA}
\definecolor{domlabel}{HTML}{1F3A93}
\begin{table*}[th]
\centering
\scriptsize
\renewcommand{\arraystretch}{1.18}
\setlength{\tabcolsep}{4pt}
\caption{\textbf{Market grounding of the SaaSBench categories (Part II).}
Continuation of Table~\ref{tab:saasbench-category-defs-c}.}
\label{tab:saasbench-category-defs-d}
\rowcolors{2}{white}{saasalt}
\resizebox{\textwidth}{!}{%
\begin{tabular}{@{}c l p{3.2cm} p{5.8cm} p{4.6cm}@{}}
\toprule
\textbf{\#} & \textbf{Dom.} & \textbf{Category} &
\textbf{Market Segment \& Rankings} &
\textbf{Representative Products} \\
\midrule
16 & CF & Accounting \& Invoicing &
\textbf{Small-Business Accounting.}
G2 Grid for Small-Business Accounting; Gartner
Critical-Capabilities for Cloud Core Financials (adjacent). &
QuickBooks Online (Intuit); Xero. \\
17 & CF & Inventory \& Warehouse &
\textbf{Inventory \& Warehouse Management.}
Gartner MQ for Warehouse Management Systems; G2 Grid for
Inventory Control. &
NetSuite Inventory Mgmt. (Oracle); Fishbowl Inventory. \\
18 & DCI & Headless CMS &
\textbf{Content Management Systems (Headless).}
Gartner MQ for Digital Experience Platforms; Forrester Wave for
Content Management Systems (Hybrid \& Headless). &
Contentful; Sanity. \\
19 & DCI & Business Intelligence &
\textbf{Analytics \& Business Intelligence Platforms.}
Gartner MQ for ABI Platforms; Forrester Wave for Augmented BI
Platforms. &
Tableau (Salesforce); Microsoft Power~BI. \\
20 & DCI & Data Catalog \& Lineage &
\textbf{Active Metadata Management / Data Catalogs.}
Gartner MQ for Metadata Management Solutions / Active Metadata;
Forrester Wave for Machine-Learning Data Catalogs. &
Alation Data Catalog; Collibra Data Intelligence Platform. \\
21 & DCI & Feature Flag \& Experimentation &
\textbf{Feature Management \& Experimentation.}
Gartner Cool Vendor recognition in Software Engineering; G2 Grid
for Feature Management. &
LaunchDarkly; Optimizely Feature Experimentation. \\
22 & DCI & Web \& Product Analytics &
\textbf{Digital \& Product Analytics.}
Gartner MQ for Digital Analytics; Forrester Wave for Digital
Intelligence Platforms. &
Google Analytics~4; Adobe Analytics. \\
23 & SI & Monitoring \& Security Ops &
\textbf{Observability / APM / Security Operations.}
Gartner MQ for Observability Platforms; Gartner MQ for SIEM
(adjacent via SecOps dashboards). &
Datadog; Splunk Observability Cloud. \\
24 & SI & Identity \& Access Management &
\textbf{Access Management / Workforce IAM.}
Gartner MQ for Access Management; Forrester Wave for Customer
Identity \& Access Management. &
Okta Workforce Identity; Microsoft Entra ID (Azure AD). \\
25 & SI & Password \& Secrets Management &
\textbf{Password Management / Enterprise Secrets.}
Gartner MQ for Privileged Access Management (adjacent); G2 Grid
for Password Manager. &
1Password Business; Bitwarden (commercial tier). \\
26 & SI & File Storage \& Sync &
\textbf{Enterprise File Sync \& Share (EFSS) / Content Collaboration.}
Gartner MQ for Content Collaboration Platforms; Forrester Wave for
Content Platforms. &
Dropbox Business; Box. \\
27 & DW & E-Signature \& Contract Management &
\textbf{Electronic Signature / Contract Lifecycle Mgmt.}
Gartner MQ for Electronic Signature; Forrester Wave for CLM. &
DocuSign; Adobe Acrobat Sign. \\
28 & DW & Low-Code / No-Code &
\textbf{Enterprise Low-Code Application Platforms (LCAP).}
Gartner MQ for Enterprise LCAP; Forrester Wave for Low-Code
Development Platforms. &
Microsoft Power Apps; OutSystems. \\
29 & DW & Electronic Health Records &
\textbf{Ambulatory / Acute EHR.}
KLAS Research EHR rankings; Gartner Hype Cycle for U.S.
Healthcare Payers \& Providers. &
Epic EpicCare; Oracle Cerner Millennium. \\
30 & DW & Workflow Automation &
\textbf{iPaaS / Integration \& Workflow Automation.}
Gartner MQ for Integration Platform as a Service (iPaaS);
Forrester Wave for iPaaS. &
Zapier; Workato. \\
\bottomrule
\end{tabular}
}
\end{table*}

\subsection{Domain Selection Principles}
\label{appendix-Domain Selection Principles}
The domain space of SaaSBench is not assembled bottom-up from an arbitrary collection of repositories. Instead, it is derived top-down from the commercial SaaS market. We begin with broad market segments and progressively refine them into categories that are distinguishable in terms of engineering patterns, until each retained category satisfies three conditions simultaneously: \textbf{(i)} it corresponds to a stable commercial use case, \textbf{(ii)} it has a recognizable product form, and \textbf{(iii)} there exists at least one production-grade open-source implementation that annotators can successfully build and launch. At the macro level, the final 30 categories are organized into 6 families along two axes: position in the value chain (front office / middle office / back office) and target service recipient (external customers vs.\ internal users or systems). This yields the six families \emph{Customer \& Growth} (CG), \emph{Productivity \& Collaboration} (PC), \emph{Commerce \& Finance} (CF), \emph{Data \& Content Infrastructure} (DCI), \emph{Security, Identity \& Infra} (SI), and \emph{Domain \& Workflow} (DW). These macro families are used only for conceptual organization and visualization. They do not alter the actual scoring dimensions of the benchmark, nor do they replace fine-grained orthogonality analysis.

Beyond market realism, we also explicitly ensure category independence through a four-dimensional orthogonality audit. For any candidate category pair \((A,B)\), we compare four dimensions in sequence: \textbf{(D1)} the core business objects, \textbf{(D2)} the core user roles, \textbf{(D3)} the dominant data read/write patterns, and \textbf{(D4)} the core architectural challenges. The decision rule is straightforward: \textbf{4/4 distinct} indicates full orthogonality; \textbf{3/4 distinct} indicates that the categories are still independent, with only one dimension exhibiting explainable local overlap; \textbf{2/4 distinct} requires additional stress testing; and \textbf{0--1/4 distinct} is treated as category overlap and must be merged or removed. This audit framework deliberately ignores generic capabilities that are shared by almost all SaaS systems, such as authentication, CRUD operations, and notifications. What truly matters is not whether two systems share boilerplate components, but whether their most distinctive engineering bottlenecks require fundamentally different technical solutions.

We apply this audit to all \(\binom{30}{2}=435\) category pairs. The final result is as follows: 421 pairs are fully orthogonal, and the remaining 14 pairs exhibit only bounded local overlap in one dimension. No category pair falls into the risk zone of 2/4 distinct or below. These 14 pairs are retained not because they are problematic but temporarily tolerated. Rather, their overlap is primarily semantic rather than engineering-essential: along the other three dimensions, especially the dimension of dominant engineering bottlenecks, they remain stably distinguishable. For example, \emph{Workflow Automation} and \emph{Low-Code / No-Code} both provide visual construction interfaces, but the former is centered on DAG execution and connector orchestration, whereas the latter is centered on UI rendering and dynamic schema binding. Similarly, \emph{Web \& Product Analytics} and \emph{Business Intelligence} both present dashboards, but the former mainly revolves around event collection and behavior telemetry under privacy constraints, whereas the latter mainly revolves around SQL / OLAP queries and chart rendering. Likewise, \emph{Identity \& Access Management} and \emph{Password \& Secrets Management} both belong to the security software stack, but the former focuses on protocol federation and session control, whereas the latter focuses on zero-knowledge secret storage and cross-device synchronization.

At the higher-level grouping layer, these six macro families satisfy the principle of \emph{mutually exclusive and collectively exhaustive} (MECE): each of the 30 categories belongs to one and only one family, while the six families together fully cover the entire task space of the benchmark without overlap. The result is a task space that is both market-grounded and low in redundancy, maximizing the diversity of core engineering primitives under a fixed benchmark budget.

\subsection{SaaS Domain Statistics}
\label{appendix-SaaS Domain Statistics}
SaaSBench contains 30 tasks in total, spanning six macro domains: \emph{Customer \& Growth} (7 tasks), \emph{Productivity \& Collaboration} (6), \emph{Commerce \& Finance} (4), \emph{Data \& Content Infrastructure} (5), \emph{Security, Identity \& Infra} (4), and \emph{Domain \& Workflow} (4). As shown in Tables~\ref{tab:saasbench-category-defs-a}, \ref{tab:saasbench-category-defs-b}, \ref{tab:saasbench-category-defs-c}, \ref{tab:saasbench-category-defs-d}, for each category we provide a brief definition, the core engineering challenge that motivates its inclusion in the benchmark, the corresponding commercial market and analyst taxonomy, and representative commercial products. Taken together, these tables make it clear that SaaSBench is not an arbitrary collection of tasks, but a benchmark grounded in a coherent commercial SaaS categorization framework.

\subsection{Task Statistics}
\label{appendix-Repo Statistics}
In Table~\ref{tab:saasbench-task-catalog}, we list the 30 final seed repositories that constitute SaaSBench. During the data curation stage, we first collect multiple candidate repositories for each category. However, under a fixed construction and annotation budget, only one seed repository is ultimately retained for each category and enters the subsequent pipeline of PRD writing, environment preparation, and evaluation construction. The retained repository must satisfy several requirements simultaneously: it should represent the typical system form of the category, exhibit signals of active maintenance, be practically buildable and deployable, and have business boundaries that are clearly aligned with the target category. Before being formally included in the benchmark, annotators also perform cold-start build validation and basic smoke testing on it. Stars denotes the approximate number of GitHub stars at the time of benchmark construction, rounded to the nearest thousand.

In addition, as shown in Table~\ref{tab:saasbench-candidate-repos}, we further release the candidate repository pool for each category. Given sufficient budget, the number of tasks in SaaSBench can be further expanded.

\definecolor{saashead}{HTML}{1F3A93}
\definecolor{saasalt}{HTML}{F4F6FA}

\renewcommand{\arraystretch}{1.25}
\setlength{\tabcolsep}{4pt}

\begin{table*}[t]
\centering
\scriptsize
\caption{\textbf{SaaSBench task catalog (30 tasks).} Each task corresponds to
one SaaS category and is instantiated from one selected open-source seed
repository. For each category, we curated multiple candidate repositories
but retained a single representative seed repository for benchmark
construction under a fixed annotation budget. We report the selected seed
repository, its approximate GitHub Stars, primary language, backend
framework, frontend stack, and main database engine. Repository URLs are
shown inline for readability; stars are rounded to the nearest thousand
at the time of benchmark construction.}
\label{tab:saasbench-task-catalog}
\resizebox{\textwidth}{!}{%
\begin{tabular}{@{}c l l r l l l l@{}}
\toprule
\textbf{\#} & \textbf{SaaS Category} & \textbf{Seed Repository} & \textbf{Stars} & \textbf{Language} & \textbf{Backend} & \textbf{Frontend} & \textbf{Database} \\
\midrule

1  & Email \& Newsletter                 & \href{https://github.com/knadh/listmonk}{listmonk}          & 19k  & Go         & Go stdlib           & Vue            & PostgreSQL          \\
2  & Customer Relationship Management    & \href{https://github.com/twentyhq/twenty}{twenty}           & 40k  & TypeScript & NestJS + GraphQL    & React          & PostgreSQL          \\
3  & Help Desk \& Ticketing              & \href{https://github.com/chatwoot/chatwoot}{chatwoot}       & 28k  & Ruby       & Rails               & Vue            & PostgreSQL          \\
4  & Form Builder \& Survey              & \href{https://github.com/formbricks/formbricks}{formbricks} & 12k  & TypeScript & Next.js API         & React (Next)   & PostgreSQL          \\
5  & Community Forum                     & \href{https://github.com/discourse/discourse}{discourse}    & 41k  & Ruby       & Rails               & Ember.js       & PostgreSQL          \\
6  & Real-time Communication             & \href{https://github.com/mattermost/mattermost}{mattermost} & 35k  & Go         & Go stdlib           & React          & PostgreSQL          \\
7  & Video Conferencing                  & \href{https://github.com/jitsi/jitsi-meet}{jitsi-meet}      & 29k  & Java / JS  & Jitsi Videobridge   & React          & N/A (Stateless SFU) \\

\midrule

8  & Project Management \& Issue Tracking & \href{https://github.com/makeplane/plane}{plane}             & 46k  & TypeScript & Django (DRF)        & React (Next)   & PostgreSQL          \\
9  & Knowledge Base \& Wiki              & \href{https://github.com/outline/outline}{outline}          & 32k  & TypeScript & Express (Node.js)   & React          & PostgreSQL          \\
10 & Time Tracking                       & \href{https://github.com/kimai/kimai}{kimai}                & 5k   & PHP        & Symfony             & SSR (Twig)     & MySQL / MariaDB     \\
11 & Gamified Productivity               & \href{https://github.com/HabitRPG/habitica}{habitica}       & 12k  & JavaScript & Express (Node.js)   & Vue            & MongoDB             \\
12 & Scheduling \& Booking               & \href{https://github.com/calcom/cal.com}{cal.com}           & 41k  & TypeScript & Next.js API + tRPC  & React (Next)   & PostgreSQL          \\
13 & Learning Management System          & \href{https://github.com/instructure/canvas-lms}{canvas-lms}& 5k   & Ruby       & Rails               & SSR            & MySQL / MariaDB     \\

\midrule

14 & E-commerce Platform
   & \href{https://github.com/medusajs/medusa}{medusa}
   & 33k  & TypeScript & Node.js / TS       & React / Next.js & PostgreSQL \\

15 & Billing \& Subscription
   & \href{https://github.com/getlago/lago}{lago}
   & 10k  & Ruby       & Rails              & React           & PostgreSQL \\

16 & Accounting \& Invoicing
   & \href{https://github.com/firefly-iii/firefly-iii}{firefly-iii}
   & 23k  & PHP        & Laravel            & SSR / Vue       & MySQL / MariaDB \\

17 & Inventory \& Warehouse              & \href{https://github.com/inventree/InvenTree}{InvenTree}    & 7k   & Python     & Django (DRF)        & React          & PostgreSQL          \\

\midrule

18 & Headless CMS
   & \href{https://github.com/payloadcms/payload}{payload}
   & 42k  & TypeScript & Next.js / Node.js   & React           & PostgreSQL \\

19 & Business Intelligence               & \href{https://github.com/apache/superset}{superset}         & 60k  & Python     & Flask / Django-style& React          & PostgreSQL          \\
20 & Data Catalog \& Lineage             & \href{https://github.com/datahub-project/datahub}{datahub}  & 12k  & Java       & Spring              & React          & MySQL / MariaDB     \\
21 & Feature Flag \& Experimentation     & \href{https://github.com/Flagsmith/flagsmith}{flagsmith}    & 6k   & Python     & Django (DRF)        & React          & PostgreSQL          \\
22 & Web \& Product Analytics            & \href{https://github.com/plausible/analytics}{plausible}    & 24k  & Elixir     & Phoenix             & React          & ClickHouse (+ PG)   \\

\midrule

23 & Monitoring \& Security Ops          & \href{https://github.com/grafana/grafana}{grafana}          & 66k  & Go         & Go stdlib           & React          & PostgreSQL          \\
24 & Identity \& Access Management       & \href{https://github.com/keycloak/keycloak}{keycloak}       & 21k  & Java       & Quarkus             & React          & PostgreSQL          \\
25 & Password \& Secrets Management      & \href{https://github.com/dani-garcia/vaultwarden}{vaultwarden}& 34k & Rust       & Actix-web           & External (BW clients) & SQLite    \\
26 & File Storage \& Sync                & \href{https://github.com/nextcloud/server}{nextcloud}       & 34k  & PHP        & Native PHP (self)   & Vue            & MySQL / MariaDB     \\

\midrule

27 & E-Signature \& Contract Management  & \href{https://github.com/docusealco/docuseal}{docuseal}     & 11k  & Ruby       & Rails               & SSR            & PostgreSQL          \\
28 & Low-Code / No-Code                  & \href{https://github.com/appsmithorg/appsmith}{appsmith}    & 39k  & TypeScript & Spring (Java) + TS  & React          & PostgreSQL (+Mongo) \\
29 & Electronic Health Records           & \href{https://github.com/openemr/openemr}{openemr}          & 5k   & PHP        & Native PHP (Laminas)& SSR            & MySQL / MariaDB     \\
30 & Workflow Automation                 & \href{https://github.com/n8n-io/n8n}{n8n}                   & 180k & TypeScript & Express (Node.js)   & Vue            & PostgreSQL          \\

\bottomrule
\end{tabular}%
}
\end{table*}
\begin{table*}[th]
\centering
\scriptsize
\renewcommand{\arraystretch}{1.08}
\setlength{\tabcolsep}{4pt}
\caption{\textbf{Candidate repository pool for the 30 SaaSBench categories.}
Each row lists the curated open-source candidate repositories for one
category; the repository ultimately selected as the benchmark seed is
shown in bold. Rows follow the benchmark task order used throughout the
paper.}
\label{tab:saasbench-candidate-repos}
\rowcolors{2}{white}{saasalt}
\resizebox{\textwidth}{!}{%
\begin{tabular}{@{}c >{\raggedright\arraybackslash}p{4.6cm} >{\raggedright\arraybackslash}p{6cm}@{}}
\toprule
\textbf{\#} & \textbf{SaaS Category} & \textbf{Candidate Repositories} \\
\midrule
1  & Email \& Newsletter &
\href{https://github.com/knadh/listmonk}{\textbf{listmonk}}; \href{https://github.com/postalserver/postal}{postal}; \href{https://github.com/mautic/mautic}{mautic} \\
2  & Customer Relationship Management &
\href{https://github.com/twentyhq/twenty}{\textbf{twenty}}; \href{https://github.com/krayin/laravel-crm}{krayin}; \href{https://github.com/salesagility/SuiteCRM}{SuiteCRM}; \href{https://github.com/espocrm/espocrm}{espocrm} \\
3  & Help Desk \& Ticketing &
\href{https://github.com/chatwoot/chatwoot}{\textbf{chatwoot}}; \href{https://github.com/uvdesk/community-skeleton}{UVdesk}; \href{https://github.com/zammad/zammad}{zammad}; \href{https://github.com/freescout-helpdesk/freescout}{freescout} \\
4  & Form Builder \& Survey &
\href{https://github.com/formbricks/formbricks}{\textbf{formbricks}}; \href{https://github.com/baptisteArno/typebot.io}{typebot}; \href{https://github.com/heyform/heyform}{heyform} \\
5  & Community Forum &
\href{https://github.com/discourse/discourse}{\textbf{discourse}}; \href{https://github.com/forem/forem}{forem}; \href{https://github.com/NodeBB/NodeBB}{NodeBB} \\
6  & Real-time Communication &
\href{https://github.com/mattermost/mattermost}{\textbf{mattermost}}; \href{https://github.com/RocketChat/Rocket.Chat}{Rocket.Chat}; \href{https://github.com/zulip/zulip}{zulip} \\
7  & Video Conferencing &
\href{https://github.com/jitsi/jitsi-meet}{\textbf{jitsi-meet}}; \href{https://github.com/bigbluebutton/bigbluebutton}{bigbluebutton}; \href{https://github.com/livekit/livekit}{livekit} \\
8  & Project Management \& Issue Tracking &
\href{https://github.com/makeplane/plane}{\textbf{plane}}; \href{https://github.com/opf/openproject}{openproject}; \href{https://github.com/Leantime/leantime}{leantime} \\
9  & Knowledge Base \& Wiki &
\href{https://github.com/outline/outline}{\textbf{outline}}; \href{https://github.com/docmost/docmost}{docmost}; \href{https://github.com/BookStackApp/BookStack}{BookStack} \\
10 & Time Tracking &
\href{https://github.com/kimai/kimai}{\textbf{kimai}}; \href{https://github.com/solidtime-io/solidtime}{solidtime} \\
11 & Gamified Productivity &
\href{https://github.com/HabitRPG/habitica}{\textbf{habitica}} \\
12 & Scheduling \& Booking &
\href{https://github.com/calcom/cal.com}{\textbf{cal.com}}; \href{https://github.com/lukevella/rallly}{rallly}; \href{https://github.com/alextselegidis/easyappointments}{easyappointments} \\
13 & Learning Management System &
\href{https://github.com/instructure/canvas-lms}{\textbf{canvas-lms}}; \href{https://github.com/openedx/openedx-platform}{openedx}; \href{https://github.com/moodle/moodle}{moodle} \\
14 & E-commerce Platform &
 \href{https://github.com/medusajs/medusa}{\textbf{medusa}}; \href{https://github.com/saleor/saleor}{saleor};\href{https://github.com/bagisto/bagisto}{bagisto}; \href{https://github.com/vendurehq/vendure}{vendure}; \href{https://github.com/Sylius/Sylius}{sylius} \\
15 & Billing \& Subscription &
\href{https://github.com/getlago/lago}{\textbf{lago}} ;\href{https://github.com/killbill/killbill}{killbill}; \\
16 & Accounting \& Invoicing &
\href{https://github.com/firefly-iii/firefly-iii}{\textbf{firefly-iii}};\href{https://github.com/akaunting/akaunting}{akaunting};  \href{https://github.com/invoiceninja/invoiceninja}{invoiceninja} \\
17 & Inventory \& Warehouse &
\href{https://github.com/inventree/InvenTree}{\textbf{InvenTree}}; \href{https://github.com/grocy/grocy}{grocy} \\
18 & Headless CMS &
\href{https://github.com/payloadcms/payload}{\textbf{payload}};\href{https://github.com/strapi/strapi}{strapi};  \href{https://github.com/directus/directus}{directus} \\
19 & Business Intelligence &
\href{https://github.com/apache/superset}{\textbf{superset}}; \href{https://github.com/metabase/metabase}{metabase}; \href{https://github.com/getredash/redash}{redash} \\
20 & Data Catalog \& Lineage &
\href{https://github.com/datahub-project/datahub}{\textbf{datahub}}; \href{https://github.com/open-metadata/OpenMetadata}{OpenMetadata}; \href{https://github.com/amundsen-io/amundsen}{amundsen} \\
21 & Feature Flag \& Experimentation &
\href{https://github.com/Flagsmith/flagsmith}{\textbf{flagsmith}}; \href{https://github.com/growthbook/growthbook}{growthbook}; \href{https://github.com/Unleash/unleash}{unleash} \\
22 & Web \& Product Analytics &
\href{https://github.com/plausible/analytics}{\textbf{plausible}}; \href{https://github.com/umami-software/umami}{umami}; \href{https://github.com/matomo-org/matomo}{matomo} \\
23 & Monitoring \& Security Ops &
\href{https://github.com/grafana/grafana}{\textbf{grafana}}; \href{https://github.com/louislam/uptime-kuma}{uptime-kuma}; \href{https://github.com/wazuh/wazuh}{wazuh}; \href{https://github.com/Graylog2/graylog2-server}{graylog}; \href{https://github.com/TwiN/gatus}{gatus}; \href{https://github.com/OneUptime/oneuptime}{oneuptime} \\
24 & Identity \& Access Management &
\href{https://github.com/keycloak/keycloak}{\textbf{keycloak}}; \href{https://github.com/goauthentik/authentik}{authentik}; \href{https://github.com/zitadel/zitadel}{zitadel}; \href{https://github.com/casdoor/casdoor}{casdoor} \\
25 & Password \& Secrets Management &
\href{https://github.com/dani-garcia/vaultwarden}{\textbf{vaultwarden}}; \href{https://github.com/bitwarden/server}{bitwarden}; \href{https://github.com/passbolt/passbolt_api}{passbolt} \\
26 & File Storage \& Sync &
\href{https://github.com/nextcloud/server}{\textbf{nextcloud}}; \href{https://github.com/haiwen/seafile}{seafile}; \href{https://github.com/owncloud/core}{owncloud} \\
27 & E-Signature \& Contract Management &
\href{https://github.com/docusealco/docuseal}{\textbf{docuseal}}; \href{https://github.com/documenso/documenso}{documenso}; \href{https://github.com/OpenSignLabs/OpenSign}{OpenSign} \\
28 & Low-Code / No-Code &
\href{https://github.com/appsmithorg/appsmith}{\textbf{appsmith}}; \href{https://github.com/nocodb/nocodb}{nocodb}; \href{https://github.com/ToolJet/ToolJet}{ToolJet}; \href{https://github.com/Budibase/budibase}{budibase}; \href{https://github.com/nocobase/nocobase}{nocobase} \\
29 & Electronic Health Records &
\href{https://github.com/openemr/openemr}{\textbf{openemr}}; \href{https://github.com/openmrs/openmrs-core}{openmrs} \\
30 & Workflow Automation &
\href{https://github.com/n8n-io/n8n}{\textbf{n8n}}; \href{https://github.com/activepieces/activepieces}{activepieces}; \href{https://github.com/automatisch/automatisch}{automatisch} \\
\bottomrule
\end{tabular}
}
\end{table*}

\section{Additional Experimental Details}
\label{appendix-additional-experimental-details}

\subsection{Inference Configuration}

For all agent-model configurations, we report \emph{pass@1}, corresponding to a single rollout per task for each configuration. We do not perform oracle-style retries after task failure. Within each rollout, the agent receives an interaction budget of up to 500 reasoning--action steps (OpenHands \emph{max\_iterations}; Codex CLI and Claude Code follow the analogous internal limit of each tool) and a wall-clock budget of $10{,}800\text{s}$ for OpenHands, Codex CLI and Claude Code. These limits cap the total time that the agent may spend on a single task. Single tool calls, such as \emph{npm install} or \emph{prisma migrate}, are allowed to run for up to $900\text{s}$ each before interruption, which empirically suffices for the most expensive installation steps in our benchmark. Network-level failures, including HTTP 429/529 errors, gateway overloads, and transient connection resets, are retried up to five times with exponential backoff between $10\text{s}$ and $60\text{s}$. Once an agent rollout terminates, either through self-declared completion or budget exhaustion, no additional retries are attempted on the same task.

\subsection{Score Aggregation}

For a single rollout, we sum the achieved scores across validation nodes and normalize the result by the task-specific \emph{total\_maxScore}, yielding a per-task score $s_t \in [0,1]$ for each task $t$. The reported \emph{benchmark score} for an agent-LLM configuration is the unweighted mean $S = \tfrac{1}{30}\sum_{t=1}^{30} s_t$, scaled to the $[0, 100]$ range. We also report a \emph{node-coverage rate}, defined as the fraction of validation nodes that reach the \emph{PASSED} state across all tasks.

\subsection{Evaluated Agents}
\label{appendix-evaluated-agents}

We evaluate three coding-agent frameworks. They differ in how they structure planning, sandbox execution, and tool calls, but all are driven by a single underlying LLM.

\paragraph{OpenHands.}
We use OpenHands~\cite{wang2024openhands} as our reference open-source agent framework and run its built-in \emph{CodeActAgent}. OpenHands is launched in local runtime mode without nested Docker, so shell tool calls of the agent execute directly inside the host shell of our orchestrator and access the task container via \emph{docker exec}. For each task-model rollout, we instantiate an OpenHands TOML configuration with the following fields: \emph{max\_iterations\,=\,500}, \emph{default\_agent\,=\,"CodeActAgent"}, an \emph{[llm]} block with \emph{temperature\,=\,0}, \emph{num\_retries\,=\,5}, \emph{retry\_min\_wait\,=\,10}, and \emph{retry\_max\_wait\,=\,60}, and a \emph{[sandbox]} block with \emph{timeout\,=\,900}. The MCP integration of the agent is disabled (\emph{enable\_mcp\,=\,false}); the agent uses only the native OpenHands tool set, including file read/write, shell, and IPython. The full reasoning trajectory of the agent is persisted to \emph{trajectory.json} for post-hoc analysis. We use OpenHands version v1.6.0.

\paragraph{Codex CLI.}
We use OpenAI Codex CLI~\cite{openai2025codexcli} through its single-shot execution mode:
\begin{promptbox}

\small
\emph{codex exec --model <id> --json --sandbox danger-full-access} \\
\emph{\hphantom{codex exec }--skip-git-repo-check -- <prompt>}

\end{promptbox}

The \emph{danger-full-access} sandbox is required because the task workspace is itself a Docker container. From the perspective of the harness, the container defines the sandbox boundary, and we therefore set the internal sandbox of Codex to behave as a no-operation layer. The \emph{--skip-git-repo-check} flag avoids spurious failures on workspaces that are not Git-initialized. The internal step budget of Codex is governed by the same wall-clock budget of $10{,}800$\,s. A thin wrapper retries at most once on transient gateway errors, including high demand, Reconnecting, and stream disconnected, with a $60$\,s backoff. Standard output, the structured JSONL event stream, and standard error are written to disk for post-hoc analysis. We use Codex CLI version codex-cli 0.128.0.

\paragraph{Claude Code.}
We use Anthropic Claude Code CLI~\cite{anthropic2025claudecodeagentic} in headless mode:
\begin{promptbox}

\small
\emph{claude -p --output-format stream-json --verbose} \\
\emph{\hphantom{claude -p }--model <id> --permission-mode bypassPermissions} \\
\emph{\hphantom{claude -p }--allowedTools <tool-list> --add-dir <workspace>} \\
\emph{\hphantom{claude -p }-- <prompt>}

\end{promptbox}
\emph{-p}/\emph{--print} runs Claude Code non-interactively. \emph{--permission-mode bypassPermissions} is the in-container analogue of \emph{danger-full-access} in Codex: it removes interactive permission prompts that would otherwise block automation. We restrict Claude Code to a fixed set of 12 core tools --- \emph{Bash}, \emph{Edit}, \emph{Write}, \emph{Read}, \emph{Glob}, \emph{Grep}, \emph{LS}, \emph{WebFetch}, \emph{NotebookEdit}, \emph{NotebookRead}, \emph{TodoRead}, and \emph{TodoWrite} --- via \emph{--allowedTools} to keep the request body small and reduce protocol-layer instability. As with Codex, transient gateway errors trigger at most one retry with a $60$\,s backoff, and the structured event stream is persisted to \emph{claude\_events.jsonl}. We use Claude Code CLI version @anthropic-ai/claude-code v2.1.126.

\paragraph{LLM Backend Versions.}
Table~\ref{tab:llm-backends} specifies the exact model snapshots used as agent backends. To support reproducibility, we record the full snapshot identifier of each vendor when available.

\begin{table*}[t]
\centering
\scriptsize
\renewcommand{\arraystretch}{1.08}
\setlength{\tabcolsep}{4pt}
\caption{LLM backends evaluated as drivers of every agent framework.}
\label{tab:llm-backends}
\begin{tabular}{@{}l l l@{}}
\toprule
\textbf{Vendor} & \textbf{Model (paper name)} & \textbf{API snapshot id} \\
\midrule
OpenAI       & GPT-5.4           & \emph{gpt-5.4-2026-03-05-xhigh} \\
Google       & Gemini 3.1 Pro    & \emph{gemini-3.1-p} \\
Anthropic    & Claude Opus 4.7   & \emph{claude-opus-4-7} \\
Moonshot AI  & Kimi K2.6         & \emph{kimi-k2-6} \\
Alibaba      & Qwen 3.6 Plus     & \emph{qwen3.6-plus} \\
DeepSeek     & DeepSeek V4 Pro     & \emph{deepseek-v4-pro} \\
Zhipu AI     & GLM 5.1           & \emph{glm-5-1-260408} \\
MiniMax      & MiniMax M2.7      & \emph{Minimax-M2.7} \\
\bottomrule
\end{tabular}
\end{table*}

we set the temperature to $0$ where supported. For reasoning models that require non-zero sampling under their thinking modes (Claude Opus, Kimi, Qwen, DeepSeek, GLM, MiniMax, and GPT-5.4 with \emph{reasoning\_effort=xhigh}), we follow each vendor's recommended \emph{temperature=1.0} and enable the corresponding thinking/reasoning channel. For Gemini, we additionally enable thought signatures (\emph{thinking.include\_thoughts\,=\,true}, \emph{thinking.budget\_tokens\,=\,8192}) and increase \emph{max\_output\_tokens} to $65{,}536$, which empirically prevents premature truncation on long tool-call chains.

\paragraph{Comparison with the Original IDE-Integrated Experience.}
We include a pragmatic note on faithfulness. All three frameworks, particularly Codex CLI and Claude Code, are routinely used inside an interactive IDE, where a human user steers the trajectory, supplies missing context, and rejects bad tool calls. Our evaluation deliberately disables this human-in-the-loop channel: the agent receives the PRD and KB once, runs autonomously inside the container until it self-declares completion or exhausts its budget, and is scored on the final container state. We expect this setting to provide a strict lower bound on the score that the same framework would achieve in real interactive use. This design isolates the autonomous engineering capability of the agent from the steering capability of the human user, which is the quantity that SaaSBench aims to measure.

\subsection{Metric Detail}
\label{appendix-metric-detail}

\paragraph{Per-node Scoring Rules.}
Let a validation node $v$ have a primitive chain $\langle p_1, p_2, \dots, p_k \rangle$ executed in order, where each primitive $p_i$ returns a Boolean success flag $\mathbb{1}[p_i \text{ passes}]$. Let $M_v \in \mathbb{R}_{>0}$ denote the pre-declared \emph{maxScore} of the node, and let $\mathrm{method}(v) \in \{\emph{binary}, \emph{weighted}, \emph{llm-as-judge}\}$ denote its scoring method. The achieved score $s_v$ is computed as follows.

\emph{(i) Binary nodes.}
These nodes are used for security-critical assertions where any failure invalidates the entire claim, such as deployment health, RBAC denials, and authentication checks:
\[
s_v^{\text{binary}} \;=\;
\begin{cases}
M_v & \text{if } \prod_{i=1}^{k} \mathbb{1}[p_i \text{ passes}] = 1, \\[3pt]
0   & \text{otherwise.}
\end{cases}
\]
For efficiency, primitive execution is short-circuited at the first failure within a binary chain.

\emph{(ii) Weighted nodes.}
These nodes are used for multi-step CRUD workflows and coverage checks where partial completion should receive partial credit:
\[
s_v^{\text{weighted}} \;=\;
\Big\lfloor\, \tfrac{1}{k} \sum_{i=1}^{k} \mathbb{1}[p_i \text{ passes}] \cdot M_v \,\Big\rfloor_{0.1},
\]
where $\lfloor \cdot \rfloor_{0.1}$ denotes rounding to one decimal place. Unlike binary nodes, weighted chains run to completion, and the achieved score scales with the fraction of primitives that pass.

\emph{(iii) LLM-as-judge nodes.}
These nodes are used only when deterministic primitives cannot adequately characterize the target, such as page-layout reasonableness or architectural-organization quality. Each such node materializes a single \emph{P17} primitive that bundles a rubric prompt, a per-node \emph{max\_score}, and an evidence source, such as the workspace codebase, the last HTTP response body, or a rendered page screenshot/HTML. The judge model is asked to return a JSON object containing a numerical \emph{score} and free-form \emph{reasoning}; we then clip the parsed score into the legal range:
\[
s_v^{\text{llm-judge}} \;=\; \mathrm{clip}\!\left(\, \mathrm{Judge}(\mathrm{rubric}_v, \mathrm{evidence}_v),\; 0,\; M_v \,\right).
\]
The judge is invoked with \emph{temperature\,=\,0} to maximize reproducibility. Concrete rubric and judge prompts are listed verbatim in Appendix~\ref{appendix-full-prompts}.

\paragraph{Status Taxonomy and Dependency Gating.}
Each node terminates in one of six statuses: \emph{PASSED} (score $> 0$ on a chain that ran cleanly), \emph{FAILED} (the chain ran but produced score 0), \emph{ERROR} (the chain raised an unhandled exception and is treated as \emph{FAILED} for scoring), \emph{SKIPPED\_DEPENDENCY} (some prerequisite of $v$ is not in \emph{PASSED}, so $v$ is not executed), \emph{SKIPPED\_LLM} (the judge API failed or LLM judging was disabled for this run), and \emph{DRY\_RUN} (an administrative skip used during evaluator authoring).

Dependency-induced skips are assigned $s_v = 0$, \emph{but their} $M_v$ \emph{remains in the denominator} of the task score. This design makes foundational failures cascade fairly: if an agent never starts the application, every dependent node is correctly counted as $0/M_v$ rather than silently dropped. At the same time, the failure is attributed to one root cause, namely the unmet prerequisite, rather than being re-charged at every downstream node, because the harness only \emph{executes} the prerequisite chain once. By contrast, \emph{SKIPPED\_LLM} nodes are removed from both the numerator and the denominator, so that a transient judge-API outage cannot artificially deflate the score of an agent. The harness logs the dropped \emph{maxScore} so that reviewers can inspect how much of the rubric pool was excluded.

\paragraph{Aggregation: Node $\to$ Task $\to$ Benchmark.}
Let $V_t$ denote the node set of task $t$, and let $V_t^{\dagger} = V_t \setminus \{v : \mathrm{status}(v) = \emph{SKIPPED\_LLM}\}$ denote the LLM-pruned set. The per-task score, scaled to $[0, 100]$, is
\[
S_t \;=\; 100 \cdot \frac{\sum_{v \in V_t^{\dagger}} s_v}{\sum_{v \in V_t^{\dagger}} M_v}\,,
\qquad
S_t^{\text{non-llm}} \;=\; 100 \cdot \frac{\sum_{v \in V_t^{\dagger}, \mathrm{method}(v) \neq \text{llm-judge}} s_v}{\sum_{v \in V_t^{\dagger}, \mathrm{method}(v) \neq \text{llm-judge}} M_v}\,,
\]
where the second formula gives the deterministic-only subscore that we additionally report for transparency. The benchmark-level score for an (agent, model) configuration is the unweighted mean across all 30 tasks,
\[
\overline{S} \;=\; \tfrac{1}{30} \sum_{t=1}^{30} S_t.
\]
We prefer the mean over a max-pooled sum because the tasks deliberately span a wide range of \emph{total\_maxScore} values, from 174 to 1{,}489 in the present benchmark. An unweighted mean of normalized per-task scores prevents oversized DAGs from dominating the headline number.

\paragraph{Per-category and Per-trajectory Diagnostics.}
In addition to the headline score $\overline{S}$, the harness emits per-category aggregates, such as \emph{Authentication}, \emph{RBAC}, \emph{Frontend}, and \emph{Deployment}, computed with the same numerator/denominator formula restricted to the relevant nodes. It also emits per-trajectory subscores for tasks that declare named user trajectories, such as \emph{happy\_path} and \emph{advanced\_workflows}.

\subsection{Primitive Taxonomy of DAG Validation Nodes}
\label{appendix-primitive-taxonomy}

The primitive chain inside each node is composed from a fixed library of primitives. Table~\ref{tab:primitives} lists the primitives that appear in at least one node across the 30 tasks, together with the number of observed node-level invocations out of $18{,}196$ primitive calls in total. Primitive identifiers are intentionally short codes (\emph{P01}--\emph{P29}, together with the browser-related \emph{RENDER\_DOM} and \emph{SCREENSHOT} primitives and the analytics-specific \emph{P\_INGEST} primitive) so that DAG JSON files remain compact and human-auditable.

\begin{table*}[th]
\centering
\scriptsize
\renewcommand{\arraystretch}{1.08}
\setlength{\tabcolsep}{4pt}

\caption{DAG validation primitives, grouped by purpose. Counts are the total number of times the primitive is invoked across the all task DAGs. One node may invoke several primitives in sequence.}
\label{tab:primitives}
\begin{tabular}{@{}l l l r@{}}
\toprule
\textbf{Group} & \textbf{Code} & \textbf{Purpose} & \textbf{Calls} \\
\midrule
\multirow{3}{*}{File / artefact}     & \texttt{P01} & file existence                              & 90 \\
                                     & \texttt{P02} & file content match (regex / substring)      & 143 \\
                                     & \texttt{P03} & file count under a path                     & 48 \\
\midrule
\multirow{4}{*}{HTTP / API}          & \texttt{P04} & HTTP request                                & 3{,}941 \\
                                     & \texttt{P05} & end-to-end CRUD round-trip                  & 47 \\
                                     & \texttt{P06} & JSON schema match                           & 123 \\
                                     & \texttt{P07} & JSON value assert (path-based)              & 2{,}595 \\
\midrule
\multirow{4}{*}{Database}            & \texttt{P08} & raw SQL / NoSQL query                       & 1{,}143 \\
                                     & \texttt{P09} & table existence                             & 533 \\
                                     & \texttt{P10} & column type / nullability check             & 506 \\
                                     & \texttt{P11} & index existence                             & 33 \\
\midrule
\multirow{3}{*}{Container / runtime} & \texttt{P12} & exec inside container                       & 438 \\
                                     & \texttt{P21} & log content check                           & 15 \\
                                     & \texttt{P23} & file upload / download                      & 9 \\
\midrule
\multirow{4}{*}{Auth / RBAC}         & \texttt{P13} & authenticated login (\textsc{Critical})     & 3{,}453 \\
                                     & \texttt{P14} & permission / role gate check                & 277 \\
                                     & \texttt{P15} & HTTP status-code assert                     & 3{,}329 \\
                                     & \texttt{P16} & response-time check                         & 4 \\
\midrule
\multirow{3}{*}{Browser / UI}        & \texttt{P18}        & browser interaction (form, click)    & 68 \\
                                     & \texttt{RENDER\_DOM}& full DOM rendering                   & 54 \\
                                     & \texttt{SCREENSHOT} & screenshot capture                   & 26 \\
\midrule
LLM judge                            & \texttt{P17} & rubric-based LLM scoring                    & 402 \\
\midrule
\multirow[c]{4}{*}{Domain-specific}  & \texttt{P\_INGEST}  & event ingestion (analytics tasks)    & 72 \\
                                     & \texttt{P19} / \texttt{P22} / \ldots & reserved / per-task         & --- \\
                                     & \texttt{P25}        & misc.\ asserts                       & 17 \\
                                     & \texttt{P27}        & webhook delivery probe               & 14 \\
\bottomrule
\end{tabular}
\end{table*}

\subsection{Six Evaluation Backbones and Category Mapping}
\label{appendix-backbone-mapping}

Each validation node in SaaSBench is annotated by the task author with a fine-grained \emph{category} string. Across the all tasks, this annotation yields 289 distinct categories, ranging from broadly applicable concepts such as \emph{DataModel} (644 nodes) and \emph{RBAC} (482 nodes) to highly task-specific concepts such as \emph{BusinessLogic\_DoubleEntry} and \emph{XMPPSignaling}. Reporting per-category scores at this level of granularity would be difficult to read and unsuitable for cross-task comparison. We therefore define a deterministic and exhaustive mapping from these 289 fine-grained categories into six high-level evaluation backbones.

\paragraph{Backbone Definitions.}
The six backbones are designed so that each captures an orthogonal axis of an enterprise SaaS system that an autonomous coding agent must implement correctly.

\begin{itemize}[leftmargin=*]
\item \emph{Deploy} --- the runnable artifact starts cleanly inside its container, including dependency installation, schema migration, environment configuration, fixture seeding, process supervision, and health checks. Source categories include \emph{Deployment}, \emph{Setup}, \emph{Build}, \emph{Configuration}, \emph{Maintenance}, \emph{Teardown}, \emph{TestFixture}, \emph{BackgroundJobs}, and \emph{CLI}.
\item \emph{Data} --- persistent state is correctly modeled and reachable, including tables, columns, indices, foreign keys, datasource wiring, file/object storage, import/export, and event-stream ingestion. Source categories include \emph{DataModel*}, \emph{Datasource}, \emph{DatabaseDiscovery}, \emph{Metadata}, \emph{Cache*}, \emph{FileSystem}, \emph{MediaManagement}, \emph{Upload}, \emph{ImportExport}, \emph{IngestionCLI}, \emph{EventProcessing}, \emph{RLS}, and \emph{Lineage}.
\item \emph{API} --- the system implements correct protocols, including RESTful CRUD, GraphQL/tRPC, WebSocket/realtime communication, outbound webhooks, search, notification fan-out, and OpenAPI conformance. Source categories include \emph{API*} and all \emph{API\_*} variants, such as \emph{API\_GraphQL}, \emph{API\_FHIR}, and \emph{API\_v2\_EE}; \emph{CRUD}/\emph{*CRUD}; \emph{GraphQL*}; \emph{tRPC}; \emph{WebSocket}; \emph{Webhook*}; \emph{Realtime}/\emph{NotificationRealtime}; \emph{Search*}; \emph{Action*}; \emph{ChatSystem}; \emph{TransactionalEmail}; and \emph{Validation}.
\item \emph{Logic} --- domain-specific business behavior built on top of the data and API layers, including orders, billing, subscription lifecycle, gamification, moderation, conferencing flows, dashboards, plugin systems, sharing/collaboration, and content templates. Source categories include all \emph{BusinessLogic*}; \emph{Workflow}; \emph{Cron*}; \emph{Async*}; the e-commerce and billing families (\emph{Order}, \emph{Cart}, \emph{Subscription}, \emph{Payment}, \emph{Tax}, \emph{Inventory}, \emph{InvoiceLifecycle}, \emph{ChargeModels}, \ldots); the gamification family (\emph{BadgeSystem}, \emph{QuestSystem}, \emph{PetMount}, \ldots); the conferencing family (\emph{Recording}, \emph{BreakoutRooms}, \emph{Whiteboard}, \emph{ConferenceFlow}, \ldots); and broad-coverage building blocks (\emph{ModerationReview}, \emph{TrustLevel}, \emph{TemplateManagement}, \emph{Plugin}, \emph{HookSystem}, \emph{Sharing}, \ldots).
\item \emph{AuthZ} --- authentication, authorization, and security auditing, covering user identity, permitted actions, and how these decisions are recorded. Source categories include \emph{Authentication}, \emph{Authorization*}, \emph{RBAC}, \emph{AccessControl}, \emph{Permission}, \emph{AuditLog}, \emph{AuditCompliance}, \emph{PasswordPolicy}, \emph{BruteForceDetection}, \emph{APIKey}, \emph{Security*}, \emph{OAuth*}, \emph{OIDCProtocol}, \emph{SAMLProtocol}, \emph{TokenExchange}, \emph{IdentityProvider}, \emph{KeyManagement}, \emph{ProtocolMappers}, \emph{UserManagement}, \emph{BusinessLogic\_2FA}, and \emph{BusinessLogic\_Identity}.
\item \emph{Quality} --- non-functional and presentation aspects that distinguish a working prototype from a production-ready system, including code architecture, frontend rendering, edge-case robustness, error handling, internationalization, and administrative user experience. Source categories include \emph{ArchitectureQuality}, \emph{Architecture}, \emph{Frontend*}, \emph{UI*}, \emph{EdgeCases}, \emph{ErrorHandling}, \emph{AdminPanel}, \emph{AdminUI}, \emph{Internationalization}, \emph{Localization}, and \emph{ConfigAndAdmin}.
\end{itemize}

\paragraph{Mapping Algorithm.}
The mapping from a fine-grained category $c$ to a backbone $B(c)$ is fully deterministic and consists of a hand-curated exact-match dictionary plus four prefix rules applied in order. We release the mapping as a single \emph{category\_to\_backbone.json} file alongside the benchmark, so any third party can re-derive the per-backbone scores from raw node-level reports. The procedure is:
\begin{promptbox}
\begin{enumerate}[leftmargin=*]
\item If $c$ is in the exact-match dictionary, return the assigned backbone.
\item Otherwise, scan the prefix rules in order and return the first match: \\ \emph{API*}/\emph{Api*} $\to$ API, \quad \emph{BusinessLogic*} $\to$ Logic, \quad \emph{DataModel*} $\to$ Data, \quad \emph{Architecture*}/\emph{Frontend*}/\emph{UI*} $\to$ Quality, \quad \emph{Auth*} $\to$ AuthZ.
\item No category falls through; the algorithm is verified to cover all 5{,}370 nodes (Table~\ref{tab:backbone-distribution}).
\end{enumerate}
\end{promptbox}

\paragraph{Resulting Distribution.}
Table~\ref{tab:backbone-distribution} shows how the 5{,}370 validation nodes and the total \emph{maxScore} of 17{,}299.1 are distributed across the six backbones under this mapping. The distribution is intentionally non-uniform: Logic dominates by maximum score (27.2\%) because business behavior is the primary factor that distinguishes a real SaaS product from a generic web application, whereas Deploy is the lightest backbone (3.7\% of \emph{maxScore}) because each task requires only a small number of nodes to certify that the service is running. We do not reweight backbones in the headline Pass@1 score. The per-backbone scores in Table~\ref{tab:main-results2} are computed independently within each backbone, so an agent that is strong on Logic but weak on Deploy is visible as such.

\begin{table*}[t]
\centering
\scriptsize
\renewcommand{\arraystretch}{1.08}
\setlength{\tabcolsep}{4pt}
\caption{Distribution of the 5{,}370 validation nodes and total \texttt{maxScore} across the six evaluation backbones, after applying the deterministic mapping in Appendix~\ref{appendix-backbone-mapping}.}
\label{tab:backbone-distribution}
\begin{tabular}{@{}l r r r r@{}}
\toprule
\textbf{Backbone} & \textbf{\#Nodes} & \textbf{\% Nodes} & \textbf{Total \texttt{maxScore}} & \textbf{\% \texttt{maxScore}} \\
\midrule
Deploy   &  289 &  5.4\% &  640.8 &  3.7\% \\
Data     &  854 & 15.9\% & 1901.0 & 11.0\% \\
API      & 1216 & 22.6\% & 3253.3 & 18.8\% \\
Logic    & 1179 & 22.0\% & 4698.0 & 27.2\% \\
AuthZ    & 1023 & 19.1\% & 3239.5 & 18.7\% \\
Quality  &  809 & 15.1\% & 3566.6 & 20.6\% \\
\midrule
\textbf{Total} & \textbf{5370} & \textbf{100.0\%} & \textbf{17299.1} & \textbf{100.0\%} \\
\bottomrule
\end{tabular}
\end{table*}

\subsection{Failure-Mode Taxonomy}
\label{app:failure-mode-taxonomy}

This appendix provides detailed definitions for the five execution-trajectory
types used in Section~\ref{Error_Analysis}. The goal of this taxonomy is
not to rank capability units by aggregate score, but to identify where the
agent's development process breaks down. We classify each capability unit using
its execution trace and node-level failure profile over six capability
backbones: deployment, data, API, business logic, authorization, and quality.
The five types are ordered from the strongest engineering trajectory (T1) to
the earliest and most severe breakdown (T5).

\paragraph{T1: Disciplined end-to-end execution.}
T1 denotes an idealized trajectory in which the agent proceeds through the
full SaaS development pipeline in a disciplined order: deployment, data schema,
authentication, business routes, authorization policy, and quality checks. At
each stage, the agent verifies the current layer before building downstream
components. A T1 unit therefore represents a stable, reproducible, and
well-validated implementation. No capability unit in our 480-unit sample falls
into this category.

\paragraph{T2: Single-backbone bottleneck.}
T2 denotes an otherwise functional system with one dominant capability
bottleneck. The stack is runnable, core data and authentication layers are
largely in place, and most business workflows are implemented, but one
capability backbone exhibits a localized failure. Examples include missing a
specific security requirement, omitting a quality constraint, or failing a
single policy-related check. This type corresponds to the common ``weakest
link'' explanation, but it accounts for only 0.6\% of our units.

\paragraph{T3: Runnable but shallow business logic.}
T3 denotes a system whose infrastructure and main execution path are working,
but whose business semantics remain incomplete. The agent can usually start the
application, create the main schema, expose endpoints, and satisfy simple happy
paths. However, it fails to operationalize detailed SaaS requirements such as
edge cases, quotas, trust levels, error handling, workflow constraints, and
policy rules. In this type, the bottleneck has moved beyond setup into
incomplete product behavior.

\paragraph{T4: Superficially reachable but structurally incomplete.}
T4 denotes a system that appears reachable from the outside but lacks reliable
foundations. For example, the HTTP entry point may return a response, while
migrations, schema constraints, authentication bootstrap, session handling, or
RBAC prerequisites remain incomplete. Downstream API and business-logic failures
then arise because they are built on an unstable base. This type captures the
case where the agent has made the project look runnable, but has not completed
the underlying engineering setup.

\paragraph{T5: Non-runnable or unstable stack.}
T5 denotes the earliest and most severe failure mode. The generated system never
becomes reliably runnable, or it becomes unstable under basic probes. Typical
causes include dependency conflicts, incorrect Docker or service configuration,
missing health checks, broken migrations, wrong working directories, port or
volume errors, and premature claims of success before the stack is actually
running. This type accounts for 63.5\% of all capability units, making it the
dominant failure mode on SaaSBench.

\section{Full Prompts}
\label{appendix-full-prompts}

We provide the full prompt templates used in SaaSBench. These templates cover two core procedures: first, prompting the agent to complete each task; second, eliciting rubric-based scores from the judge model at \texttt{llm-as-judge} validation nodes. All templates follow the exact concatenation logic used in our public release.

\subsection{Agent Task Prompt}
\label{appendix-agent-prompt}

For each (task, agent, model) configuration, the harness constructs a single textual prompt by concatenating four blocks. Square-bracketed placeholders are populated on a per-task basis, while all other text remains unchanged across the 30 tasks.

\begin{templatebox}{Block 1 --- anti-cheat banner}
\textbf{Mandatory anti-cheat policy.} 
You MUST implement the platform from scratch within this Docker environment. Cloning, copying, or otherwise importing any pre-existing open-source codebase (via \texttt{git clone}, \texttt{wget}, \texttt{curl}, container image extraction, package downloads of unrelated projects, etc.) is strictly forbidden and will be detected by the harness. Trajectories that fetch external source repositories receive a score of 0 regardless of the resulting test outcomes.
\end{templatebox}

\begin{templatebox}{Block 2 --- role and environment specification}
You are a senior full-stack engineer. Your task is to build a complete \texttt{[task-specific product description]} from scratch inside an already-running Docker environment.

\medskip

\textbf{Environment.} The Docker environment is already up and running --- you do not need to pull images or start containers. You work inside the container \texttt{[container-name]}, with \texttt{/app} as the working directory (currently empty). Run commands inside the container like this: \verb|docker exec [container-name] bash -c "your command"|.

\medskip

\textbf{Tools pre-installed in the container} (Ubuntu 22.04): \texttt{[per-task list of language runtimes, package managers, native libraries, database clients]}.

\medskip

\textbf{Database} (\texttt{[engine name and version]}, already running): host \texttt{db}, port \texttt{[port]}, database \texttt{[db]}, username \texttt{[user]}, password \texttt{[pwd]}, extensions \texttt{[ext-list]}.

\medskip

\textbf{The application MUST listen on port \texttt{[port]}.}

\medskip

\textbf{What you need to do}: (1) create a complete \texttt{[stack name]} project inside \texttt{/app}; (2) install dependencies; (3) run database migrations; (4) create the following 3 evaluation users (admin / moderator / regular user) with the credentials specified in the PRD; (5) start the application server, listening on \texttt{0.0.0.0:[port]}.

\medskip

\textbf{Key technical requirements.} \texttt{[per-task framework version, ORM choice, cache, job queue, auth header conventions, etc.]}
\end{templatebox}

\begin{templatebox}{Block 3 --- requirements document and KB pointer}
\#\#\# Requirements Document\\
\\
The complete product requirements document has been placed in the working directory at \texttt{/app/task.md}. Read it carefully and implement everything it describes.\\
\\
In addition, the working directory contains a supplementary knowledge base at \texttt{/app/knowledge\_base.json} that clarifies ambiguous points in the PRD. Each entry has the shape \verb|{id, question, answer, source_reference, confidence}|. Consult it as a reference when \texttt{task.md} is unclear on details such as field semantics, state transitions, serialization, filtering, or permissions. You can read it on demand with \texttt{jq}, \texttt{cat}, or \texttt{grep} (e.g.\ \verb|jq '.clarifications[] | {id, question, answer}' /app/knowledge\_base.json|); you do not have to read the whole file at once.
\end{templatebox}

\begin{templatebox}{Block 4 --- success condition}
\#\#\# Success Condition\\
\\
The application server must be reachable on the designated port by the time you consider the task done. Anything that needs to happen before that --- dependency install, schema migration, fixture creation, process supervision --- is part of what is being evaluated; the harness will not enumerate the steps for you.
\end{templatebox}

The PRD and KB are not embedded directly in the prompt body. Instead, they are placed in the workspace as files, namely \texttt{/app/task.md} and \texttt{/app/knowledge\_base.json}. This design allows the agent to re-read them on demand without unnecessarily expanding the rolling context window. Since PRDs in SaaSBench contain 4{,}363 lines on average, embedding them in every model turn would otherwise consume the entire context budget.

\subsection{LLM-as-Judge Rubric Prompt}
\label{appendix-judge-prompt}

For each \texttt{llm-as-judge} node, the \texttt{P17} primitive collects a node-specific rubric together with a piece of evidence, such as a workspace codebase listing, the last HTTP response body, or a rendered page screenshot/HTML. It then issues a single chat-completion call to the judge model, Claude Sonnet 4.5, with temperature $=0$. The two-message template is defined as follows.

\begin{templatebox}{System message}
You are an expert evaluator. Score the evidence against the rubric. Respond ONLY with a JSON object: \verb|{"score": <0-{max_score}>, "reasoning": "<brief explanation>"}|
\end{templatebox}

\begin{templatebox}{User message}
\#\# Rubric\\
\{rubric\_prompt\}\\
\\
\#\# Evidence\\
\{evidence\}
\end{templatebox}

The reply from the judge is parsed as a JSON object. The integer \texttt{score} is clipped into $[0, M_v]$ to guard against rare cases in which the judge ignores the requested upper bound. Markdown code fences in the reply are stripped before \texttt{json.loads} is applied. If parsing fails or the upstream API is unavailable, the node is marked as \texttt{SKIPPED\_LLM} and excluded from both the numerator and the denominator of the per-task score, as detailed in Appendix~\ref{appendix-metric-detail}.

\paragraph{Concrete rubric example.}
For illustration, we include the actual rubric used for the \texttt{FE\_HOMEPAGE\_LAYOUT} node of the Community Forum task. The task identifier is \texttt{task\_aoiwqoiq}, the maximum score is 6, and the evidence source is the rendered homepage screenshot together with HTML.

\begin{templatebox}{Concrete rubric example}
Evaluate the homepage layout quality: (1) Does a topic list display with titles, authors, timestamps? (2) Is there a navigation bar with logo, search, login/signup? (3) Is there a category sidebar or navigation? (4) Are visual hierarchy and spacing professional? Score 0--6.
\end{templatebox}

The remaining 401 \texttt{P17} invocations across the 30 tasks follow the same rubric structure: a short numbered checklist of 3--6 observable criteria paired with an explicit score range. Together, these rubrics form a consistent and audit-friendly contract between the benchmark authors and the judge model.

\section{Simplified Task Example}
This section uses a concrete task, Discourse, to illustrate two core artifacts of a SaaSBench instance. The following two simplified snippets show how a single SaaSBench task is instantiated in practice.

\begin{templatebox}{Simplified PRD}
\#\#\# Task: Build a Full-Featured Community Forum Platform\\
\\
\textbf{Product Background.}
The task asks the agent to build a production-ready community discussion
platform for large-scale, structured conversations. The system supports
threaded topics, posts, categories, tags, badges, private messages,
notifications, full-text search, moderation workflows, and an extensible
plugin architecture. It also includes an administrative dashboard for site
settings, users, groups, email, backups, themes, API keys, webhooks, reports,
and operational monitoring.\\
\\
\textbf{Technical Stack.}
The required backend stack is Ruby~3.4 with Ruby on Rails~8.0, PostgreSQL~13+,
ActiveRecord, Redis, Sidekiq, mini\_scheduler, and MessageBus-based real-time
updates. The frontend is implemented with Ember.js and Glimmer components,
using pnpm and Node.js~20+. Search is implemented with PostgreSQL full-text
search, and file storage must support local storage or S3-compatible object
storage.\\
\\
\textbf{API Conventions.}
The platform does not use a global \texttt{/api/} prefix. JSON responses are
served through the same Rails controllers as HTML, selected by either a
\texttt{.json} suffix or the \texttt{Accept: application/json} header.
Authentication supports browser sessions, global API keys through
\texttt{Api-Key} and \texttt{Api-Username} headers, and per-user API keys
through \texttt{User-Api-Key}. Cookie-authenticated mutating requests require
CSRF protection, and clients obtain a token from \texttt{GET /session/csrf}.
Errors follow a structured JSON shape with \texttt{errors} and
\texttt{error\_type}.\\
\\
\textbf{Data Model.}
The PRD specifies a large PostgreSQL schema with over one hundred domain
entities, including users, posts, topics, categories, groups, tags, badges,
notifications, uploads, bookmarks, invites, reviewables, API keys, drafts,
topic timers, post actions, topic users, group memberships, user emails,
post revisions, themes, color schemes, webhooks, polls, search data, user
security keys, and many administrative or audit-related tables. The schema
uses ActiveRecord conventions, foreign keys, indexed lookup columns,
soft-deletion fields, polymorphic associations, and domain-specific enums.\\
\\
\textbf{Core Business Workflows.}
The forum implements a progressive trust-level system from TL0 to TL4.
Trust levels control posting, private messages, uploads, likes, edits,
moderation-like privileges, and topic management. Promotions are automatic
for TL1--TL3 based on reading activity, visits, posts, replies, likes, and
flag history, while TL4 is manually granted. The authorization layer is modeled
as a per-request Guardian permission system that checks category visibility,
topic creation, post editing, post deletion, staff actions, category
moderation, and user management permissions.\\
\\
\textbf{Post and Topic Lifecycle.}
Post creation is handled by a service pipeline that validates spam rules,
slow mode, rate limits, PM recipient limits, and category permissions before
creating posts and topics. Raw Markdown is converted into sanitized HTML
through a markdown-it and server-side sanitization pipeline. New topics and
posts update user statistics, topic counters, tracking states, search data,
and MessageBus channels. Topic lifecycle behavior includes closing, archiving,
pinning, unlisting, private messages, topic timers, auto-close behavior,
post moving, revisions, and whisper posts for staff-only discussions.\\
\\
\textbf{Moderation, Search, and Notifications.}
The system includes flag queues, reviewable objects, review scores, threshold
actions, silence and suspension workflows, watched words, spam handling,
screened emails/IPs/URLs, user warnings, and staff action logs. Notifications
cover replies, mentions, quotes, likes, private messages, high-priority events,
muting rules, watched topics, and first-post watching. Search supports
PostgreSQL full-text search, ranked results, category and tag filters, user
filters, status filters, date filters, and advanced query operators.\\
\\
\textbf{Frontend Requirements.}
The Ember.js frontend must expose topic list views, category-scoped and
tag-scoped listings, topic detail pages, user profiles, user preferences,
admin dashboard pages, search pages, review queues, group pages, tag pages,
private-message views, badges, bookmark manager, static content pages,
safe-mode diagnostics, setup wizard, authentication pages, and error or
permission-boundary pages. Interactive components include the topic composer,
Markdown preview, uploads, emoji picker, mentions, hashtag autocomplete,
notification menus, admin editors, filters, pagination, and responsive layouts.\\
\\
\textbf{User and Permission System.}
Authentication includes browser session cookies, global API keys, user API
keys, OAuth/OmniAuth, SSO, passkeys, two-factor authentication, and passwordless
email login. Roles include admin, moderator, staff, and trust-level users.
Permissions combine role checks, trust levels, category permissions, ownership,
staff status, group membership, rate limits, and visibility rules. Restricted
categories and private messages require explicit user or group authorization.\\
\\
\textbf{Deployment and Runtime Contract.}
The generated system must run in the provided Docker environment with Rails,
PostgreSQL, Redis, Sidekiq, Node.js, pnpm, and the Ember build pipeline. The
application must expose the configured HTTP port, initialize the database,
run migrations and seed data, start background workers, compile frontend
assets, and provide a health-checkable web service.
\end{templatebox}

\begin{templatebox}{Simplified local DAG fragment example}
\#\#\# Core Forum Workflow Fragment\\
\\
\textbf{Node 1: \texttt{DEPLOY\_HEALTH}.}
Objective: verify that the generated forum application is reachable on
\texttt{localhost:8020}. Actions: send \texttt{GET /}. Validations: the returned
status code belongs to the accepted healthy-status set.\\
\\
\textbf{Node 2: \texttt{AUTH\_SESSION\_LOGIN}.}
Prerequisites: \texttt{AUTH\_CSRF\_TOKEN}, \texttt{AUTH\_CREATE\_ADMIN}.
Objective: verify browser-session login. Actions: fetch a CSRF token from
\texttt{/session/csrf.json}, then submit administrator credentials to
\texttt{/session}. Validations: the response status is 200 and the returned
user is \texttt{eval\_admin}.\\
\\
\textbf{Node 3: \texttt{CRUD\_TOPIC\_CREATE}.}
Prerequisite: \texttt{AUTH\_CREATE\_ADMIN}. Objective: verify topic creation.
Actions: submit \texttt{POST /posts.json} with a title, body, and category.
Validations: the response contains a \texttt{topic\_id}, the first post has
\texttt{post\_number = 1}, and the topic row exists in PostgreSQL.\\
\\
\textbf{Node 4: \texttt{CRUD\_POST\_CREATE}.}
Prerequisite: \texttt{CRUD\_TOPIC\_CREATE}. Objective: verify reply creation.
Actions: create a reply under the newly created topic. Validations: the returned
post has \texttt{post\_number = 2}, and the topic's \texttt{posts\_count} is
updated to 2 in the database.\\
\\
\textbf{Node 5: \texttt{CRUD\_CATEGORY\_CREATE}.}
Prerequisite: \texttt{AUTH\_CREATE\_ADMIN}. Objective: verify category
creation. Actions: submit \texttt{POST /categories.json} with category name and
colors. Validations: the response returns the created category name, and the
category row exists in PostgreSQL.\\
\\
\textbf{Node 6: \texttt{SEARCH\_BASIC}.}
Prerequisite: \texttt{CRUD\_TOPIC\_CREATE}. Objective: verify full-text search.
Actions: create a topic containing the unique keyword
\texttt{SearchXyz42}, wait for indexing, and query
\texttt{/search.json?q=SearchXyz42}. Validations: search results exist and the
result blurb contains the keyword.\\
\\
\textbf{Dependency structure.}
\[
\texttt{AUTH\_CREATE\_ADMIN}
\rightarrow
\texttt{CRUD\_TOPIC\_CREATE}
\rightarrow
\{\texttt{CRUD\_POST\_CREATE},\ \texttt{SEARCH\_BASIC}\}
\]
\[
\texttt{AUTH\_CSRF\_TOKEN},\ \texttt{AUTH\_CREATE\_ADMIN}
\rightarrow
\texttt{AUTH\_SESSION\_LOGIN}
\]
\[
\texttt{AUTH\_CREATE\_ADMIN}
\rightarrow
\texttt{CRUD\_CATEGORY\_CREATE}
\]
\end{templatebox}

\section{Broader Impact}
\label{appendix:broader-impact}
SaaSBench aims to support more realistic and diagnostic evaluation of long-horizon coding agents. By focusing on enterprise-level SaaS development scenarios, the benchmark can help researchers better understand whether current agents can move beyond localized code generation toward end-to-end system construction, deployment, and validation. We hope that SaaSBench can promote the development of coding agents that are more reliable, transparent, and better aligned with practical software engineering needs.

At the same time, more capable coding agents may lower the barrier to software creation in both beneficial and harmful ways. They can improve developer productivity, broaden access to software development, and help non-expert users rapidly prototype useful systems. However, they may also generate insecure code, propagate hidden defects, or be misused to automate harmful software behavior.

\section{Limitations and Future Work}
\label{appendix:limitations}

\paragraph{Limitations.}
Real-world software development is highly diverse and continues to evolve with changes in organizational contexts, deployment practices, engineering conventions, and product requirements. Therefore, SaaSBench may not cover all possible variants of enterprise software construction, nor can it exhaust all qualitative factors involved in engineering decisions. These factors do not diminish the value of the benchmark. Rather, they indicate that SaaSBench still has room for further refinement as coding agents and software engineering workflows continue to develop.

\paragraph{Future Work.}
Future work can further extend SaaSBench in two directions. First, it can increase the number of tasks, expand the set of SaaS categories, and cover a broader range of technology stacks to improve the representativeness of the benchmark. Second, future versions can support more dynamic software development scenarios, such as iterative requirement updates, system maintenance tasks, and multi-stage product evolution.

%%%%%%%%%%%%%%%%%%%%%%%%%%%%%%%%%%%%%%%%%%%%%%%%%%%%%%%%%%%%

\newpage

\end{document}